\documentclass[useAMS,usenatbib,usegraphicx]{mn2e}

\newcommand{\araa}{ARA\&A}   \newcommand{\aap}{A\&A}
\newcommand{\aj}{AJ}         \newcommand{\apj}{ApJ}
\newcommand{\apjl}{ApJ}      \newcommand{\apjs}{ApJS}
\newcommand{\mnras}{MNRAS}

\title[Evolution of the {\it u}-band luminosity function]{Evolution of the {\it u}-band luminosity function from redshift 1.2 to 0}
\author[M. Prescott, I. K. Baldry and P. A. James]{Matthew Prescott\thanks{E-mail:
mxp@astro.livjm.ac.uk}, Ivan K. Baldry and Phil A. James\\Astrophysics Research Institute, Liverpool John Moores University, Twelve Quays House, Egerton Wharf, Birkenhead, CH41 1LD}
\date{Accepted 2009 April 01. Submitted in original form 2008 November 13}

\begin{document}

\pagerange{\pageref{firstpage}--\pageref{lastpage}} \pubyear{2009}

\maketitle

\label{firstpage}

\begin{abstract} 
We produce and analyse {\it u}-band ($\lambda \approx 355$ nm) luminosity
functions for the red and blue populations of galaxies using data from the
Sloan Digital Sky Survey (SDSS) {\it u}-band Galaxy Survey ({\it u}GS) and
Deep Evolutionary Exploratory Probe 2 (DEEP2) survey. From a spectroscopic
sample of 41\,575 SDSS {\it u}GS galaxies and 24\,561 DEEP2 galaxies, we
produce colour magnitude diagrams and make use of the colour bimodality of
galaxies to separate red and blue populations. Luminosity functions for eight
redshift slices in the range $0.01 < z < 1.2$ are determined using the
$1/V_{max}$ method and fitted with Schechter functions showing that there is
significant evolution in $M^{*}$, with a brightening of 1.4 mags for the
combined population. The integration of the Schechter functions yields the
evolution in the {\it u}-band luminosity density out to $z\sim1$. By
parametrizing the evolution as $\rho \propto (1+z)^{\beta}$, we find that
$\beta = 1.36 \pm 0.2$ for the combined populations and $\beta = 2.09 \pm
0.2$ for the blue population. By removing the contribution of the old stellar
population to the {\it u}-band luminosity density and correcting for dust
attenuation, we estimate the evolution in the star formation rate of the
Universe to be $\beta_{SFR} = 2.5 \pm 0.3$. Discrepancies between our result
and higher evolution rates measured using the infrared and far-UV can be
reconciled by considering possibilities such as an underestimated dust
correction at high redshifts or evolution in the stellar initial mass
function.

\end{abstract}


\section{Introduction}
\label{intro}

The evolution in the star formation rate (SFR) of the Universe over
the course of cosmic history is a measure of great importance that has
many applications in various fields of astrophysics. In addition to
the obvious applications to the build-up of stellar mass in galaxies,
other examples include constraining the stellar initial mass function
(IMF) \citep{Baldry03,Wilkins08}, testing models of chemical evolution
in the Universe \citep{Calura04}, constraints on the nature of both
type Ia supernovae \citep{Strolger04} and gamma-ray bursts
\citep{Dainge06, Li08} and comparisons with global AGN activity and
the growth of black holes \citep{Franceschini99, Somerville08}.

SFRs of galaxies have been measured by a variety of methods, at
different wavelengths. These include: SFRs determined from the far- and
mid-infrared produced by the re-emission of ultraviolet (UV) photons
by dust grains, revealing hidden star formation
\citep{LeFloch05,Per-Gon05}; SFRs measured from radio
emission produced by relativistic electrons in supernova remnants
\citep{Condon92,Seymour08}; Nebular emission lines such as ${\rm
  H_{\alpha}}$, ${\rm H_{\beta}}$ and ${\rm [OII]}$, produced by the
recombination of ionized gas surrounding hot young OB stars
\citep{Kennicutt83,Moustakas06,Cooper08,James08a}; and X-ray emission
produced by X-ray binary systems in late-type galaxies
\citep{Norman04,Lehmer08}.

More directly, SFRs have been measured from the UV luminosities of
galaxies produced from short lived OB stars (lifetimes $\sim 10^{7}$
yr). Early studies such as \citet{Lilly96} and \citet{Madau96} used
this to establish that the volume-averaged SFR of the Universe has
been in decline since $z \sim 1$ to the present day. Although later
studies \citep{Madau98, Steidel99, Wilson02} measured the decline with
improved precision, it is still poorly constrained and its cause is
still a major problem in the field of observational cosmology.

As an alternative to the far UV luminosity, the {\it u}-band ($\lambda \approx
355$ nm) luminosity of a galaxy, predominantly emitted by young stars of ages
$< 1$ Gyr, has been shown to be a reasonable tracer of star formation
\citep{Hopkins03}. Using {\it u}-band has two main advantages over the far UV,
because i) more data are available to improve the statistics and ii) it is less
affected by extinction. \citet{Driver08} find that at $z = 0$, the fraction of
photons to escape through the dust from the local galaxy population is $\sim
45$ per cent in the {\it u}-band as opposed to $\sim 25$ per cent at
200-nm. One disadvantage of the {\it u}-band as a tracer of star formation is
the contribution from old stellar populations, which is dominant in 
early-type galaxies and small in star forming galaxies. 
This has to be accounted for.

So far {\it u}-band field luminosity functions (LFs) have been produced by
\citet{Blanton03} for Data Release 2 (DR2) of the Sloan Digital Sky Survey
(SDSS) main sample, by \citet{Baldry05} for the SDSS {\it u}-band Galaxy
Survey ({\it u}GS) and more recently by \citet{Dorta08} using DR6, probing
volumes out to $z \sim 0.2$. In this paper we extend the analysis to higher
redshift, by combining samples of galaxies taken from the SDSS {\it u}GS and
the Deep Evolutionary Exploratory Probe 2 (DEEP2) survey, to produce LFs in
the range $0.01 < z < 1.2$, which represents $\sim 8$ Gyr of cosmic history
from when the Universe was $\sim 5$ Gyr old. Selected by {\it R}-band, the
DEEP2 survey is ideally suited for comparison to the SDSS {\it u}GS as {\it
  R}-band magnitudes at $z \sim 1$ correspond to rest {\it u}-band
magnitudes. We go on to produce colour magnitude diagrams (CMDs) from the data
and make use of the observed colour bimodality \citep{strateva01, Baldry04} to
separate the narrow band of red, passive galaxies and blue star forming
populations of galaxies. Integration of the LFs produced from these separate
populations allows the luminosity density, or total output of {\it u}-band
light in a given volume to be determined.
       
The structure of this paper is as follows: Section 2 describes the DEEP2
Survey and SDSS {\it u}GS, and the data reduction employed to obtain our final
galaxy samples. In section 3 the methods required to calculate the luminosity
functions including completeness corrections and K corrections are described,
and we present rest-frame CMDs displaying how the red and blue populations are
separated. In section 4 we present the luminosity functions, fitted Schechter
parameters and luminosity densities. Our results are then compared with the
observations at different wavelengths and we discuss our estimate of the
evolution of SFR of the Universe in section 5. Section 6 contains a summary of
our conclusions.  Throughout this paper we adopt a cosmology with parameters
$(\Omega_{m}, \Omega_{\Lambda})_{0} = (0.3, 0.7)$ and $H_{0} = 70$ km s$^{-1}$
Mpc$^{-1}$. Magnitudes are corrected for Galactic extinction using dust maps
of \citet{Schlegel98}.
   
\section{Data}
\subsection{DEEP2}

The Deep Evolutionary Exploratory Probe 2 (DEEP2) survey
\citep{Davis03} is a redshift survey primarily designed to probe the
properties of galaxies at redshift z $\sim$ 1. When fully complete the
spectra of around 50\,000 galaxies with redshifts $z < 1.4$ 
will have been measured over an area of sky totalling 3.5 deg$^2$.

In this paper we use data from the third data release, DR3, which
comprises two catalogues: photometric and spectroscopic. The
photometric catalogue contains the $B, R$ and $I$ magnitudes of
716\,465 objects (including stars and duplicate objects) obtained with
the $12k \times 8k$ mosaic camera \citep{Cuillandre01} on the 3.6 m
Canada-France-Hawaii-Telescope. The spectroscopic catalogue contains
redshifts and spectra of 46\,337 objects measured by the DEIMOS
spectrograph \citep{Faber03} with wavelength range $600-920$ nm and
resolution $\lambda /\Delta \lambda \sim 5000$ on the 10m Keck 2
telescope. Galaxies selected for redshift measurements are limited to
apparent AB magnitudes in the range $18.5 < R < 24.1$.
   
The survey area is separated into 4 regions of the sky and is made up
of a 0.5 sq. deg. field known as the Extended Groth Strip (EGS) and
three 1.0 sq. deg. fields designated fields 2, 3 and 4. Galaxies in
fields 2, 3 and 4 are preselected by their $B$, $R$ and $I$ magnitudes
using colour cuts, to have estimated redshifts $z > 0.7$. No such
preselection was carried out for the EGS which means that there is an
approximately equal number of galaxies below and above $z =0.7$ in
this region.

\subsection{DEEP2 sample reduction} 

Before determining the luminosity functions, stars and duplicate
objects had to be removed from the DEEP2 catalogues and it was also
useful to cut the area of the photometric catalogue, due to the
spectroscopic catalogue not yet covering the same area as the
photometric catalogue. The methods used to reduce the catalogues to
obtain a final sample of galaxies used in this paper are described in
this section.
   
The first reduction of the photometric catalogue was to cut down the
areas of the photometric survey in each field, so that the areas of
the photometric and incomplete spectroscopic catalogues are the
same. 177\,680 objects in the photometric catalogue were cut, reducing
it to 538\,785.

The second procedure was to discard galaxies with magnitudes outside
the range $ 18.5 < R < 24.1$, to give the spectroscopic and
photometric catalogues equal magnitude limits. This removed 313\,362
objects from the photometric catalogue and 58 objects from the
spectroscopic catalogue.  After matching up the spectroscopic object
numbers with those in the photometric catalogue, 6 objects were found
to have no corresponding object number in the photometric
catalogue. These 6 objects were removed from the spectroscopic
catalogue. After this reduction the photometric and spectroscopic
catalogues contained 225\,423 and 46\,279 objects respectively.

The third reduction was to remove stars from the catalogue. Objects in
the photometric catalogue are designated a quantity, PGAL
\citep{Coil04}, that gives the probability of an observed object being
a galaxy or not. This probability is determined from star-galaxy
separations calculated from the magnitudes, sizes and colours of the
objects. Resolved sources i.e. those which are definitely galaxies are
given a PGAL value of 3 whereas unresolved sources are given a
probability of being a galaxy between 0 and 1. In this investigation
it was decided to use objects with PGAL $> 0.2$ as in
\citet{Willmer06}.
 
The final reduction was to remove galaxies in the catalogues that had been
observed multiple times and designated different object numbers. These
duplicated observations for objects were identified by running a procedure to
find and match objects within 0.5 arc-seconds of each other. The matched
objects with the best redshift quality and smallest error in $R$ mag, were
used to produce the luminosity functions. The duplicate objects with poorer
data were then removed from the photometric and spectroscopic catalogues.
These two further processes reduced the photometric catalogue by 58\,824 to
166\,599 galaxies. 629 objects in the spectroscopic catalogue which had object
numbers corresponding to galaxies removed from the photometric catalogue, were
also removed, leaving 45\,650 objects in the spectroscopic catalogue.
              
The properties of the reduced fields can be seen in Table
\ref{table1}. $N{_P}$ and $N{_z}$ are the numbers of galaxies with photometry
and spectroscopy. Our primary sample consists of 24\,561 galaxies with
redshifts in the range $0.4 < z < 1.2$.

\begin{table}
\caption{Properties of the DEEP2 reduced survey sample}
\label{table1}
\begin{center}
\begin{tabular}{lccc} \hline
Field                       &Reduced Area (sq. deg.)  &$N{_P}$ &$N{_z}$\\
\hline 
EGS                         &0.45  &36319  &13440\\
Field 2                     &0.68  &34976  &9559\\
Field 3                     &1.00  &52390  &11904\\
Field 4                     &0.94  &42914  &10747\\
\hline
Total                       &3.07 &166559  &45650\\                         
\hline
\end{tabular}
\end{center}
\end{table}    

\subsection{The SDSS {\it u}-band Galaxy Survey}

The Sloan Digital Sky Survey is only briefly described here, for
greater detail the reader is referred to \citet{York00} and subsequent
data release papers such as the EDR paper \citep{Stoughton02}. The
SDSS is both a photometric and spectroscopic survey which makes use of
a dedicated 2.5-m telescope situated on Apache Point, New
Mexico. Photometry in 5 broadband filters, {\it ugriz}, is conducted
using a mosaic CCD camera \citep{Gunn98} and calibrated with a 0.5-m
telescope \citep{Hogg01}. Spectra of selected objects are measured
using a 640-fibre fed spectrograph with wavelength range 380-920 nm
and resolution $\lambda / \Delta \lambda \sim 1800$.

The sample of galaxies used in this paper is taken from `Stripe 82' of
the SDSS `Southern Survey' \citep{SDSSDR4}. This consists of a
$2.52^{\circ}$ wide strip across the Southern Galactic Pole (SGP)
centred along the celestial equator from RA $-50.8^{\circ}$ to
$58.6^{\circ}$. Multiple imaging passes of the stripe allow co-added
$u$-band Petrosian fluxes to be determined.  This gives an increase in
the S/N ratio by a factor of about 3 over single epoch observations.
Extra spectroscopic observations of galaxies (fully described in
\citealt{Baldry05}) include a low-$z$ $r_{petro}<19.5$ sample and a
$u_{select}<20$ sample.\footnote{$u_{select} = u_{model} - r_{model} +
  r_{petro}$ and can be regarded as pseudo {\it u}-band Petrosian
  magnitude. This selection magnitude was used because the
  spectroscopic observations were targeted on single epoch data, where
  $u_{petro}$ is less reliable at these faint limits.}  The multiple
scans, and extra spectra, allow a catalogue of objects to be produced
with $u_{petro} < 20.5$ based on the co-added data.  The result is a
photometric catalogue of 75\,087 objects, with magnitudes in the range
$14.5 < u < 20.5$ and spectroscopy for 48\,640 objects in an area of
275 deg$^2$. Our primary sample consists of 41\,615 galaxies with
redshifts in the range $0.01 < z < 0.2$.

\section{Methods}

\subsection{Completeness corrections}

Completeness corrections are required to correct the luminosity
functions for the number of galaxies that are not targeted for
spectroscopic observations. For both surveys, empirical completeness
values were determined by binning in observed quantities. For the SDSS
{\it u}GS sample of galaxies it is assumed that all the redshifts are
correct but for the DEEP2 data we additionally take into account the
redshift success rate, which is explained below.

\subsection{DEEP2 completeness}

Redshifts in the DEEP2 survey were first determined by an automated
pipeline (Cooper et al. 2009, in preparation) which does all the image
reduction and extraction of spectra which are then validated by
eye. In this process the redshifts were given a quality assessment
value (Q) from 1 to 4. Q=1 indicates the redshift determined to be
poor due to a lack of features in the spectra, whereas Q=4 means the
redshift has two or more identified features (the ${\rm [OII]}$
$\lambda =372.7$ nm doublet is considered to be two features) and the
determined redshift is rock solid.

\begin{figure}
\includegraphics[width=0.47\textwidth]{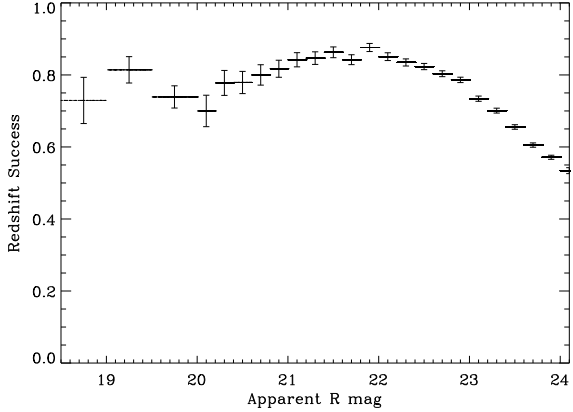}
\caption{Redshift success fraction as a function of apparent $R$-band magnitude for DEEP2 DR3. The redshift success fraction is simply the number of galaxies with redshift quality Q $\ge$ 3 divided by the total number of galaxies, within an apparent $R$-band magnitude bin. Error bars indicate Poisson errors.} 
\label{RedSucess}
\end{figure}

In this paper we only use galaxies with Q $\ge$ 3. To calculate the
redshift success, galaxies are divided into bins of $R$ band apparent
magnitude from $18.5 < R < 24.1$. Brighter galaxies with $18.5 < R <
20.0$ are divided into 0.5 mag bins whereas galaxies with $20.0 < R <
24.1$ are divided into 0.2 mag bins. The redshift success is then
simply the number of galaxies with Q $\ge$ 3 divided by the total
number of galaxies per magnitude bin. The redshift success as a
function of $R$ band apparent magnitude is shown in
Fig.~\ref{RedSucess}.

On inspection of Fig.~\ref{RedSucess} it is clear that the redshift success
drops unexpectedly for the brightest galaxies. After correspondence with Mike
Cooper of the DEEP2 collaboration, we have found this is mainly due to some
low redshift galaxies that have been incorrectly designated as having too poor
quality spectra (to measure a redshift). The DEEP2 team acknowledge this is
something that will be corrected for the next data release (DR4).  Our
completeness corrections assume that the missed redshifts, both bright and
faint, have a similar redshift distribution as the obtained redshifts.  The
overall conclusions of this paper are unlikely to be significantly affected by
the validity of this assumption because the completeness is $>0.7$ up to
$R\simeq23$.

A targeting completeness correction accounts for the galaxies that are
not targeted for spectroscopy. This is calculated by dividing galaxies
in both photometric and spectroscopic catalogues into colour-colour
bins. Firstly the galaxies are divided into 39 $R-I$ colour bins
between $-3.0 < R-I < 4.0$. Galaxies are divided in the following bins
$[< -0.03, -0.03\rightarrow 1.01 $ (binsize 0.04), $1.01\rightarrow
  1.56 $ (0.05), $ > 1.56]$.  Then the galaxies are subdivided into 61
$B-R$ colour bins. All galaxies lie within the range $-1.6 < B-R <
7.0$ and are divided into bins $[< -1.0, -1.0\rightarrow 4.9 $
  (binsize 0.1), $ > 4.9]$. This is done with the intention of
obtaining a few tens of galaxies in each colour-colour bin.

\begin{figure}
\includegraphics[width=0.47\textwidth]{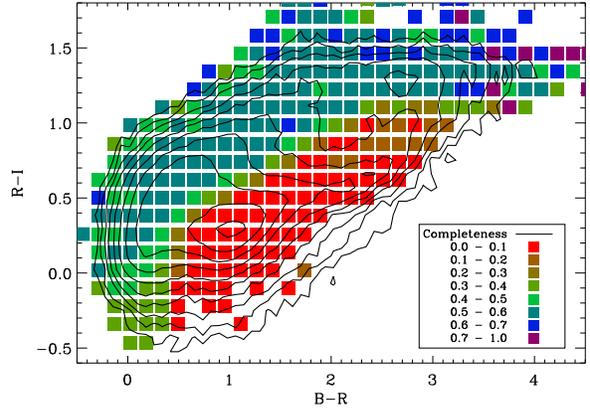}
\caption{Targeting completeness of DEEP2 galaxies as a function of $R-I$ and
  $B-R$ colours. The targeting completeness corrects the LFs for galaxies not
  selected for spectroscopic observations. It is the number of galaxies in the
  spectroscopic catalogue divided by the number of galaxies in the photometric
  catalogue, within a colour-colour bin. All four DEEP2 fields are treated the
  same way regarding targeting completeness. A consequence of this is two
  distinct regions; a region of moderate completeness consisting of galaxies
  from fields~2, 3 and~4, having preselected colours to have redshifts greater
  than 0.7 and a region of low completeness made up of galaxies from the
  Extended Groth Strip. The distribution of colours for 166\,599 DEEP2
    galaxies from the photometric catalogue are represented by 
    logarithmically-spaced contours.}
\label{CCplot}
\end{figure}

Targeting completeness is simply the number of galaxies from the
spectroscopic catalogue divided by the number of galaxies in the
photometric catalogue in each colour-colour bin. Fig.~\ref{CCplot} shows the distribution of colours for all 166\,599
  DEEP2 galaxies in the photometric catalogue represented as contours
  along with the targeting completeness as a function of $R-I$ and
  $B-R$. From this plot the colour preselection used to obtain
  galaxies at $z > 0.7$ can be clearly seen as the region of moderate
  completeness (green). In this paper we choose not to treat galaxies
  within the Extended Groth Strip (EGS) separately from the other
  DEEP2 fields and an effect of this is the region of galaxies with
  low completeness (red) also seen in the figure.

\begin{figure}
\includegraphics[width=0.47\textwidth]{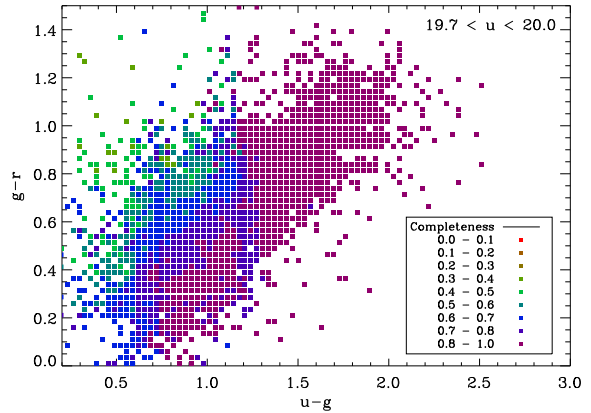}
\caption{Sample completeness as a function of $g-r$ and $u-g$ colours for the SDSS {\it u}GS galaxies with $19.7 < u < 20.0$. For completeness as a function of $\Delta_{sg}$, {\it u}, $u-g$ and $g-r$ see \citet{Baldry05}.}
\label{SDSSCCplot}
\end{figure}

The overall completeness for a single galaxy is simply the product of the redshift success rate and targeting completeness. 

\subsection{SDSS {\it u}GS completeness}

Completeness corrections for the SDSS {\it u}GS sample are fully described in
\citet{Baldry05}. These are determined by dividing the sample into bins using
the star-galaxy separation, $\Delta_{sg}$ ($= r_{PSF} - r_{model}$, the
difference between {\it r}-band point spread function and `model' magnitudes),
along with {\it u}, $u-g$ and $g-r$ as variables. Firstly the objects are
divided into weakly and strongly resolved sources at the dividing line of
${\Delta}_{sg} = 0.3$. For $u > 19.7$ strongly resolved sources are divided
into 0.1 mag bins and weakly resolved sources are divided into 0.2 mag
bins. These magnitude bins are then further divided on the basis of $u-g$ and
$g-r$ colour bins so that there are around 50 galaxies in each bin. The
completeness is simply the fraction of photometric objects that have
spectroscopy. The completeness as a function of $g-r$ and $u-g$ colours for
galaxies with $19.7 < u < 20.0$ is presented in Fig.~\ref{SDSSCCplot}. For a
more detailed figure showing the completeness as a function of $\Delta_{sg}$,
{\it u}, $u-g$ and $g-r$ see Fig.~6 of \citet{Baldry05}. Although the
  completeness can be low ($\sim$ 0.1--0.2) for the faintest blue galaxies
  with $20.3 < u < 20.5$, all parts of galaxy colour-colour space over the
  full magnitude range are sampled.

\begin{figure*}
\includegraphics[width=0.40\textwidth]{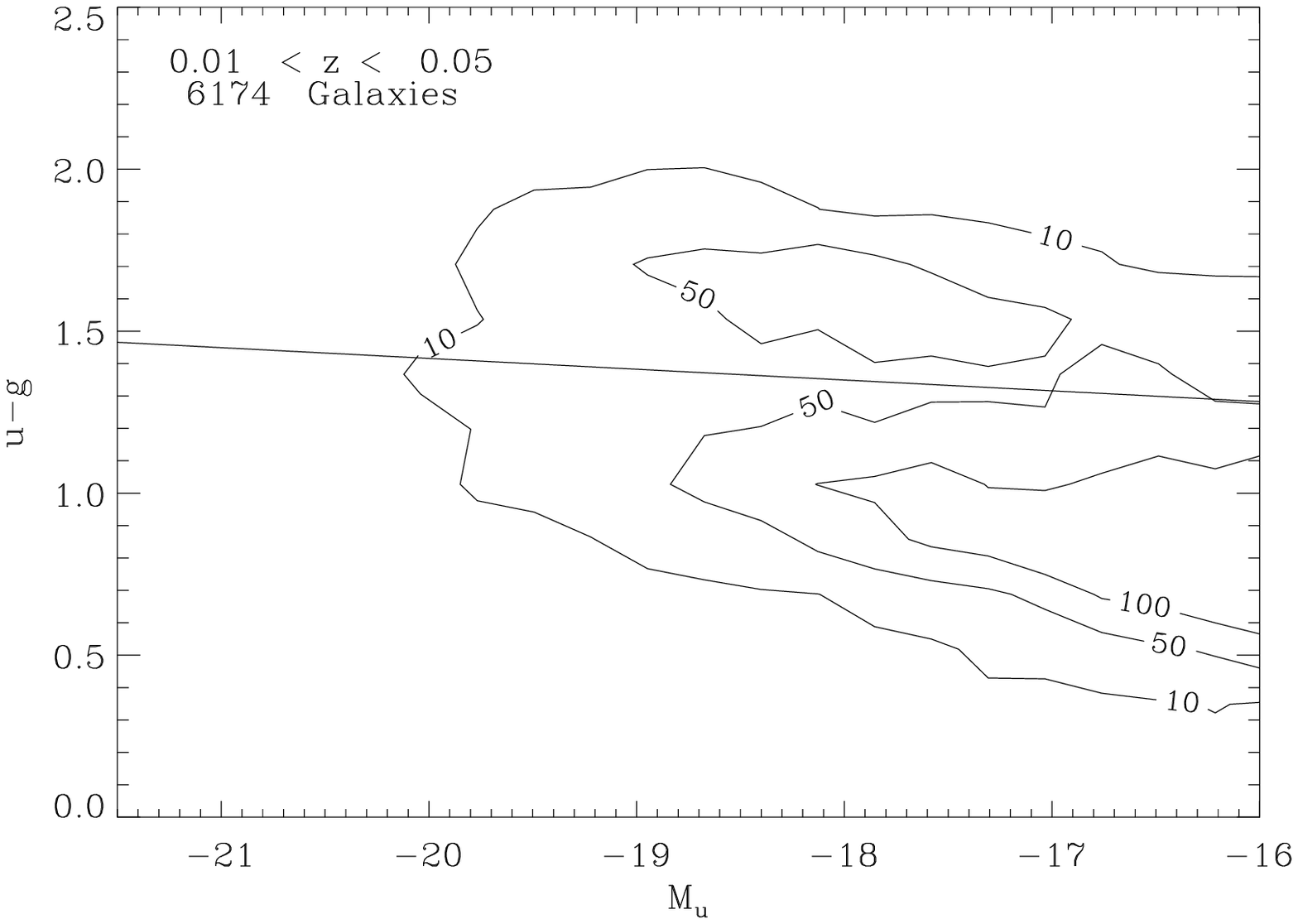}
\includegraphics[width=0.40\textwidth]{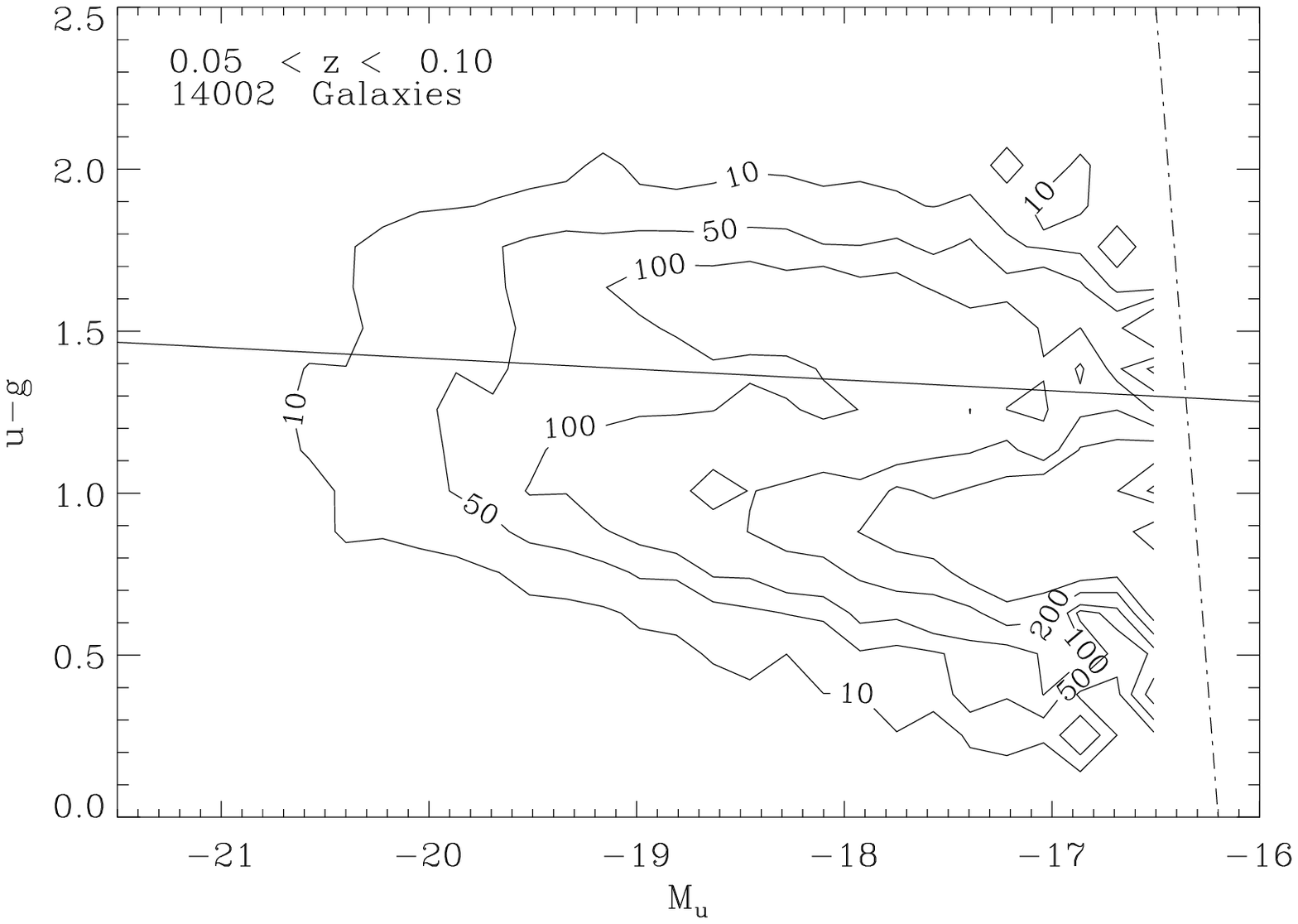}
\includegraphics[width=0.40\textwidth]{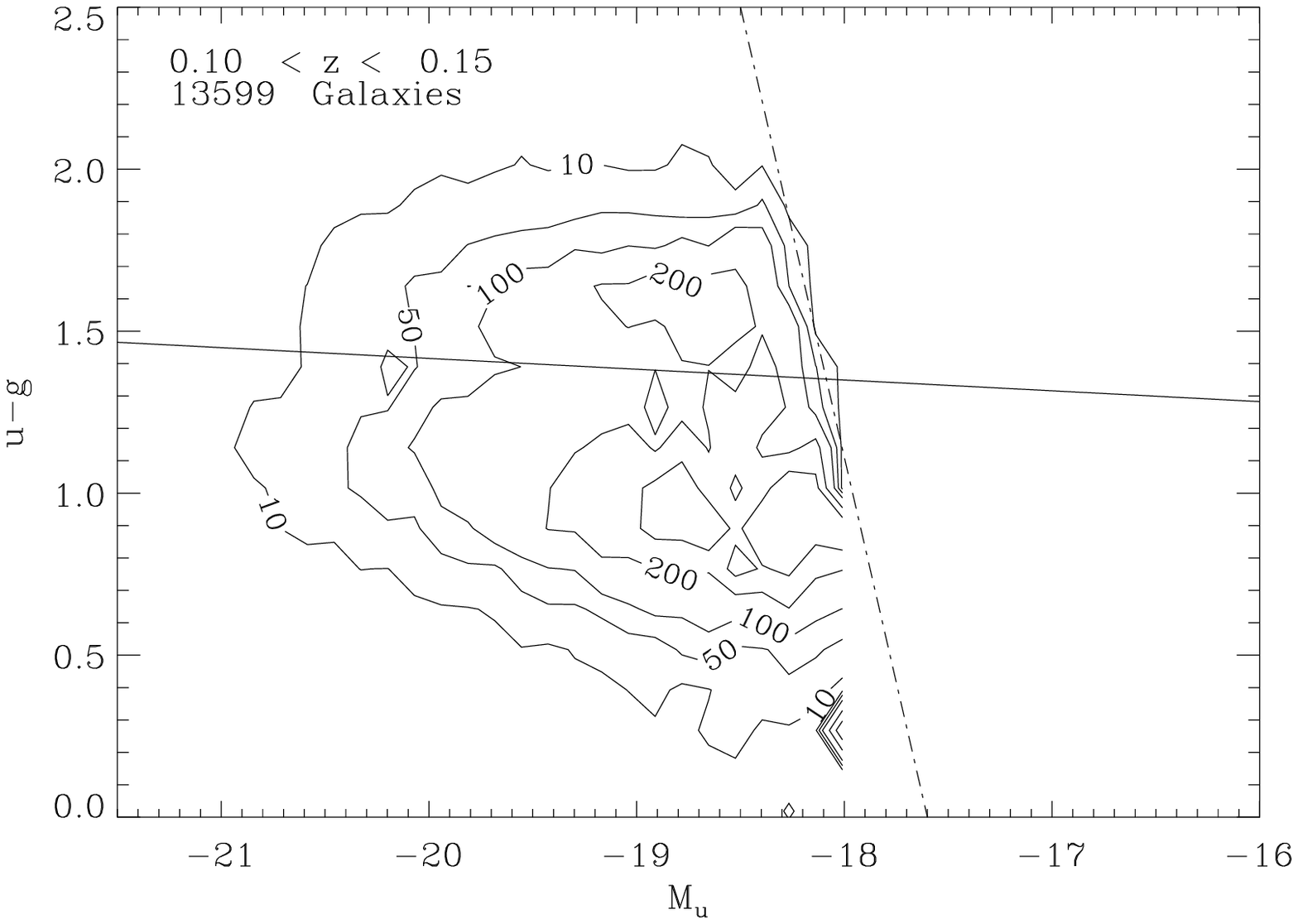}
\includegraphics[width=0.40\textwidth]{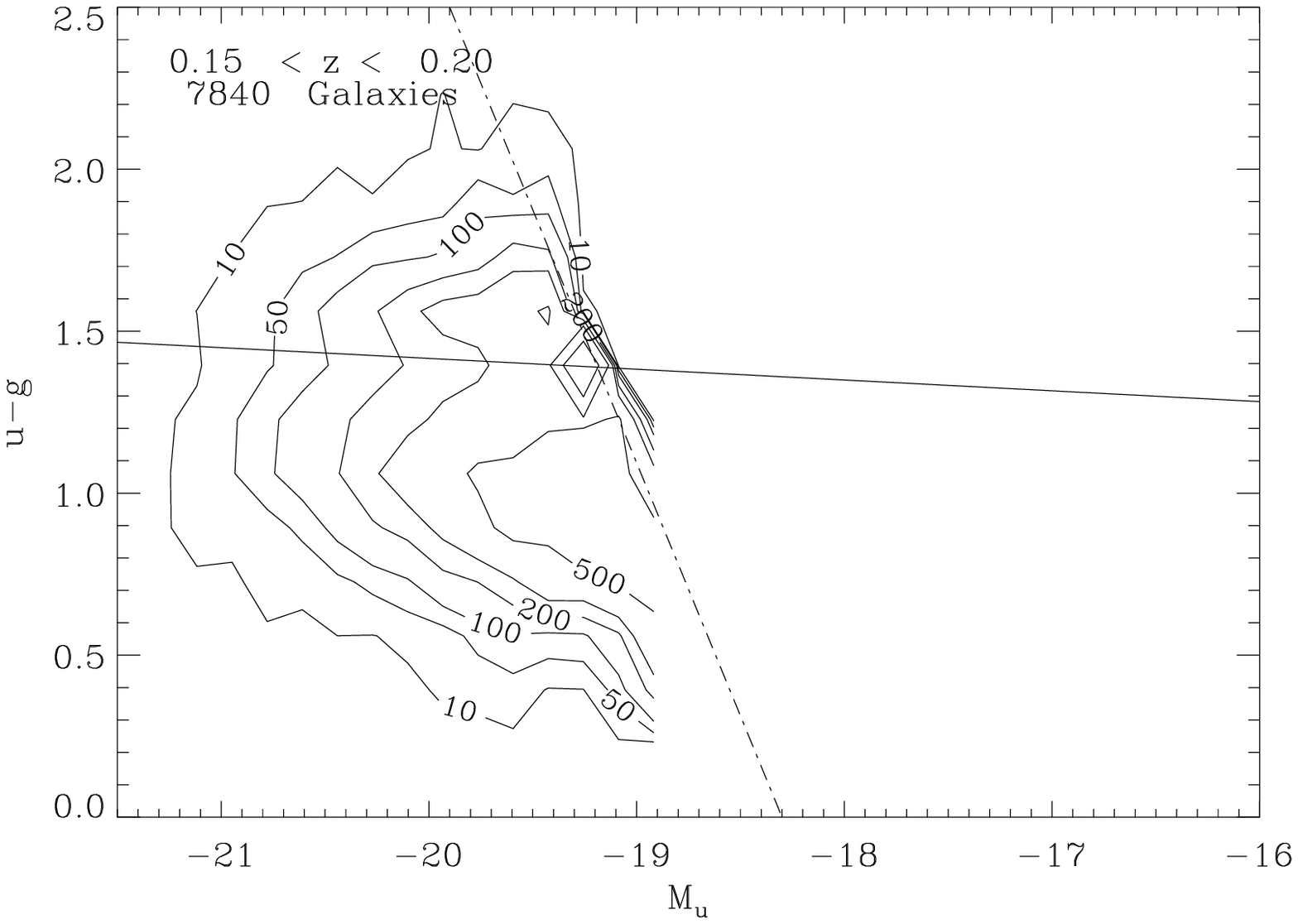}
\caption{Colour magnitude diagrams for the four SDSS redshift slices, in the range $0.01 < z < 0.2$. The number densities are weighted by $1/(V_{max}C_{i})$ and are represented by logarithmic contours. Little evolution is seen in the colour corresponding to the region between red and blue galaxies for all redshift slices. The solid line represents the straight line given by eq.1, that is used to divide the red and blue galaxy populations. The dashed lines represent the magnitude limit of the redshift slice.}
\label{SDSScont}
\vspace*{0.2cm}
\includegraphics[width=0.40\textwidth]{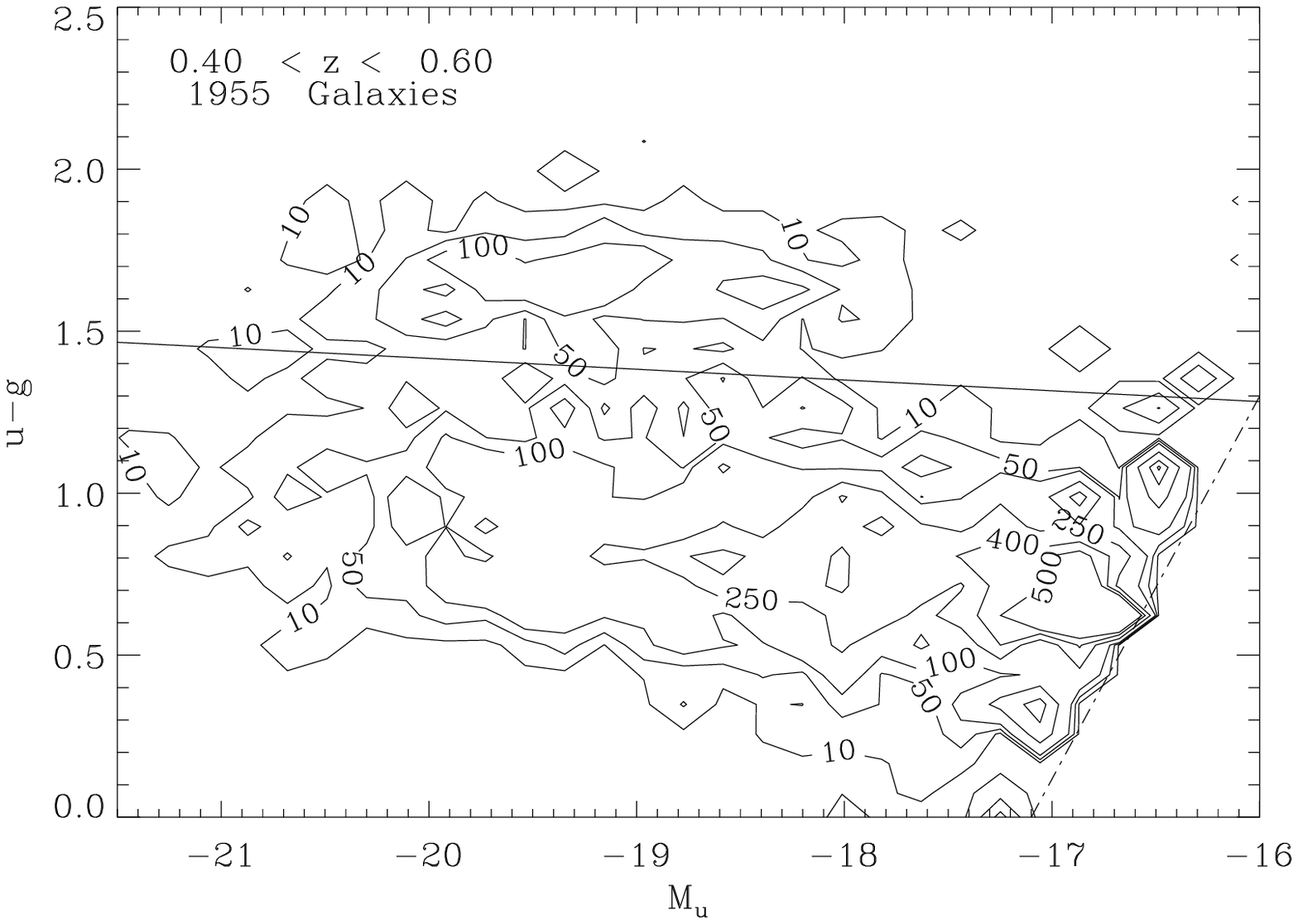}
\includegraphics[width=0.40\textwidth]{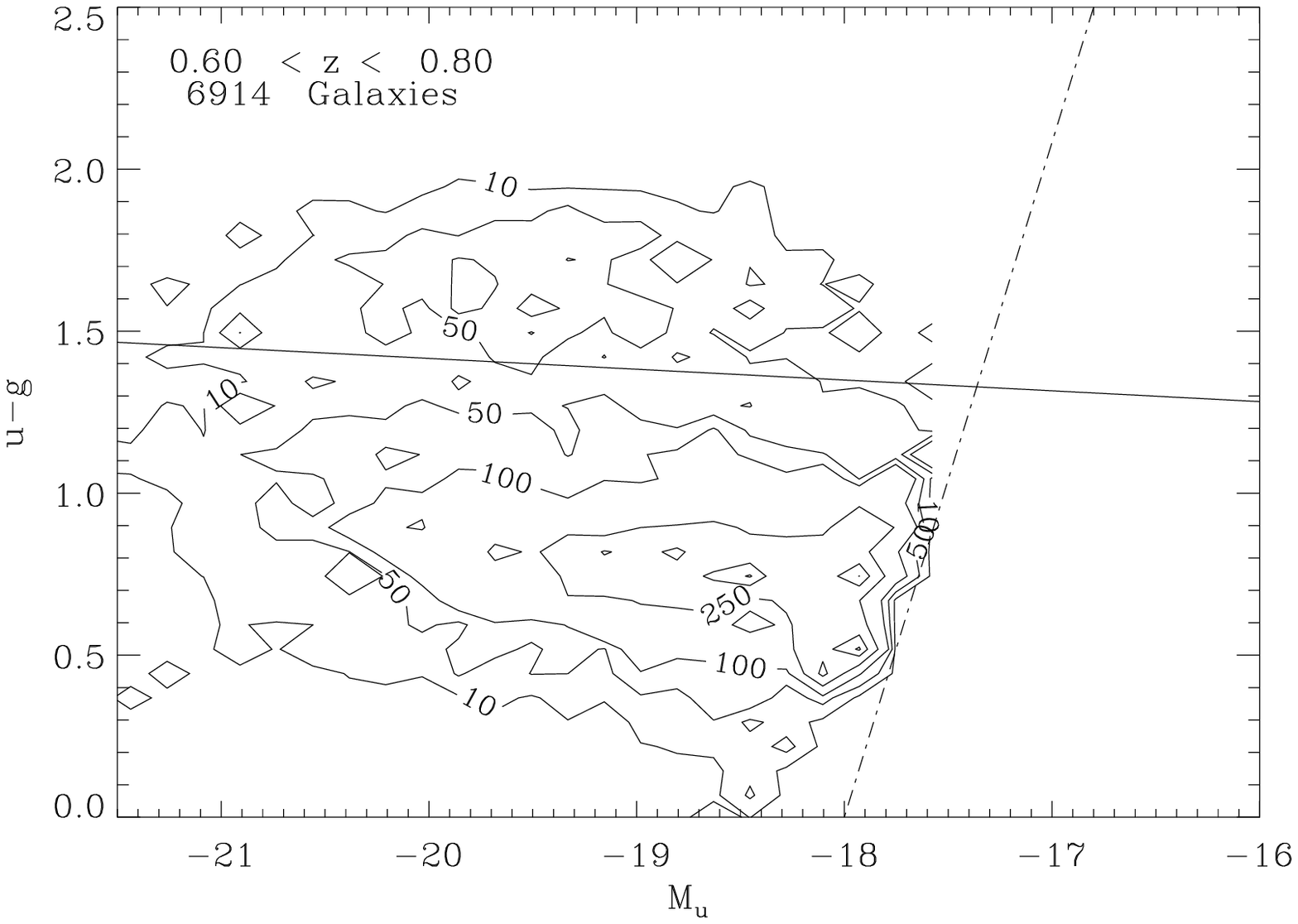}
\includegraphics[width=0.40\textwidth]{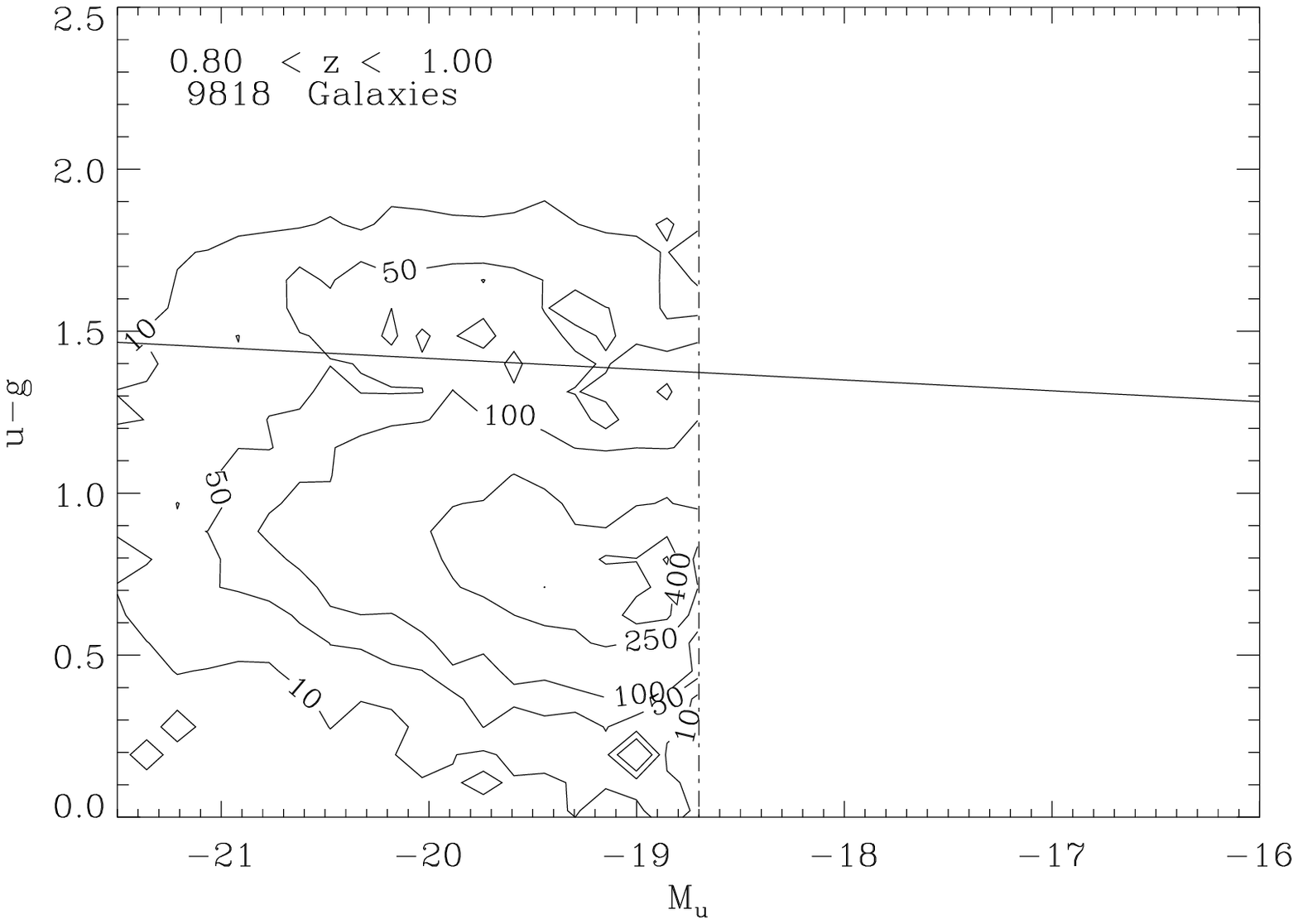}
\includegraphics[width=0.40\textwidth]{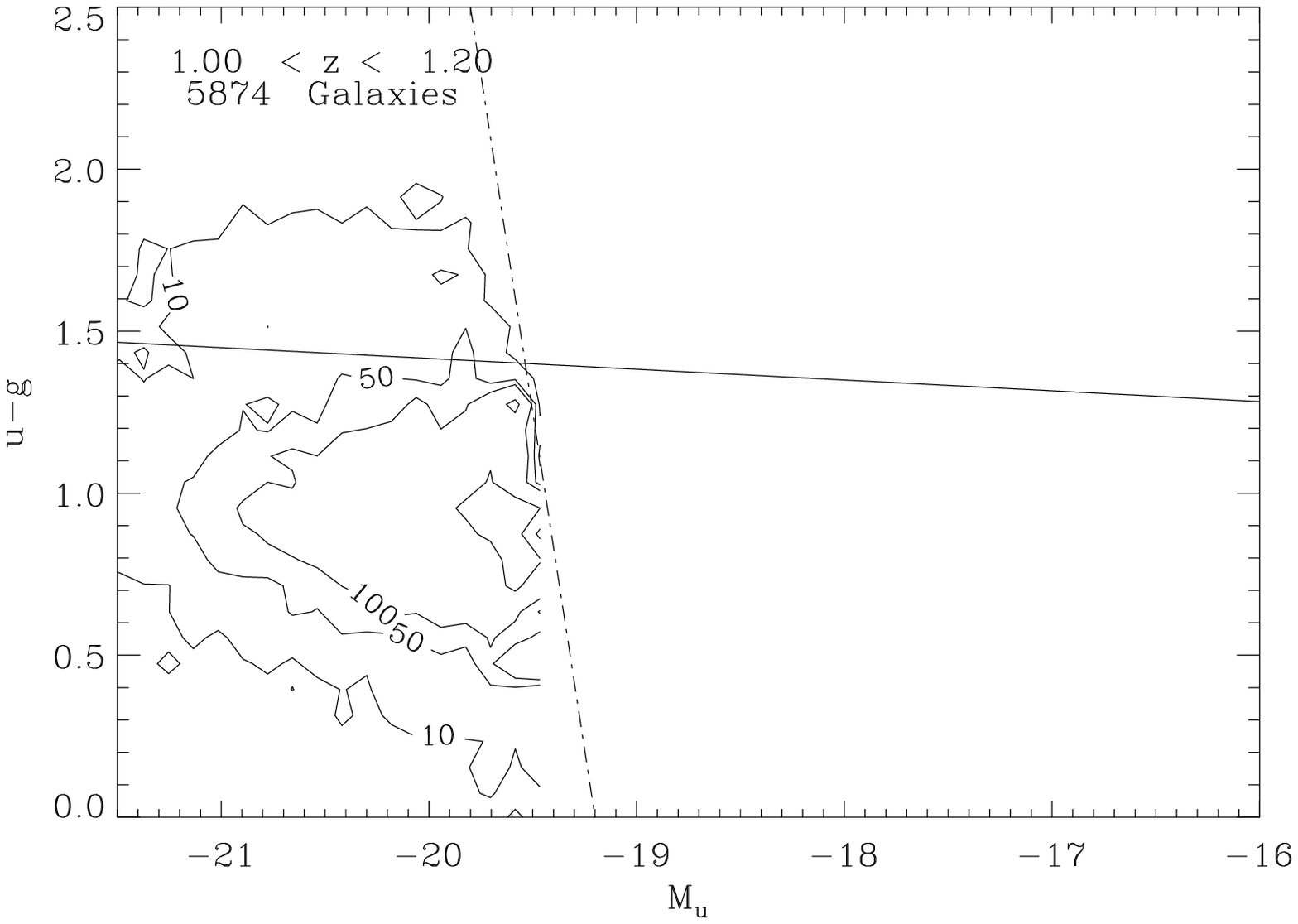}
\caption{CMD contour plots for four DEEP2 redshift slices in the range $0.4 < z < 1.2$. Number densities are represented as logarithmic contours. The solid line represents the straight line given by eq.1 that is used to divide the red and blue galaxy populations. The dashed lines represent the magnitude limit of the redshift slice.}
\label{DEEP2cont}
\end{figure*}
          
\subsection{K corrections and colour magnitude diagrams}

To convert the observed apparent magnitudes of the galaxies to
absolute rest-frame magnitudes we make use of the K correction program
Kcorrect (version 4.1.4) of \citet{Blanton03K}. For the DEEP2 sample,
$B,R$ and $I$ magnitudes (effective wavelengths $\approx$ 439, 660,
813.5 nm) are transformed into rest frame SDSS {\it u} and {\it g}
magnitudes (effective wavelengths 355 and 467 nm).

For the SDSS {\it u}GS sample, the observed {\it u} and {\it g}
magnitudes are converted into the rest frame. This process enabled the
production of rest frame {\it u-g} colour magnitude diagrams (CMDs)
for the galaxies in 8 different redshift slices. Fig.~\ref{SDSScont}
shows CMDs produced from SDSS data and Fig.~\ref{DEEP2cont} shows CMDs
from DEEP2. These are represented as contour plots which are weighted
by $1/(V_{max}C_{i})$. Here $V_{max}$ is the maximum comoving volume,
within which a galaxy could lie within a redshift slice and within the
magnitude limits of the survey \citep{Schmidt68} and $C_{i}$ is the
completeness.  These plots show the build up in red galaxies over time
and that the colour bimodality of galaxies extends out to $z\sim1$, as
found by \citet{Bell04}, \citet{Willmer06}, \citet{Cirasuolo07},
\citet{Franzetti07} and \citet{Cowie08}.
       
On inspection of the DEEP2 and SDSS {\it u}GS contour plots, we find
although both blue and red populations are brightening in {\it u}-band
magnitude as redshift increases, there is no obvious evolution in
$u-g$ colours for the natural dividing line. This lack of evolution
has also been observed by \citet{Cowie08} in AB3400-AB4500 colours out
to $z =1.5$ and \citet{Willmer06} in $U-B$ colours. Given this lack of
change, red and blue galaxies at all redshifts can be separated by a
straight line with the equation:
\begin{equation}
u-g = -0.033M_{u}+0.75
\end{equation}
Changing this dividing line will have little effect on the overall results, as long as it remains in the gap between blue and red populations.

\begin{figure}
\includegraphics[width=0.47\textwidth]{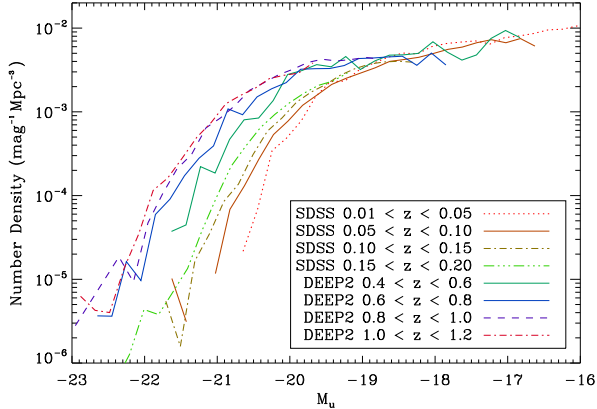}
\caption{Evolution of the {\it u}-band luminosity function for the combined red and blue populations of galaxies. LFs are produced for four redshift slices between $0.01 < z < 0.2$ for the SDSS {\it u}GS data and four redshift slices between $0.4 < z < 1.2$ for DEEP2, using the non-parametric $1/V_{max}$ method. Galaxies are binned in 0.2 mag bins.}
\label{LFcomp}
\end{figure}

\begin{figure}
\includegraphics[width=0.47\textwidth]{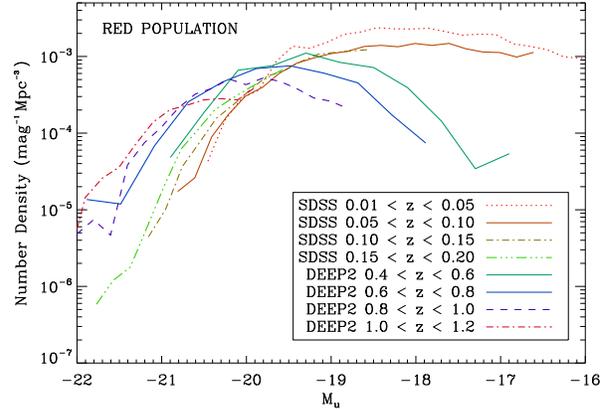}
\caption{Evolution of the {\it u}-band luminosity function for the red population of galaxies. 0.2 mag bins are used to produce all the LFs apart from the $0.4 < z < 0.6$ and $0.6 < z < 0.8$ DEEP2 redshift slices where 0.4 mag binning is used.}  
\label{LFcompred}
\end{figure}

\begin{figure}
\includegraphics[width=0.47\textwidth]{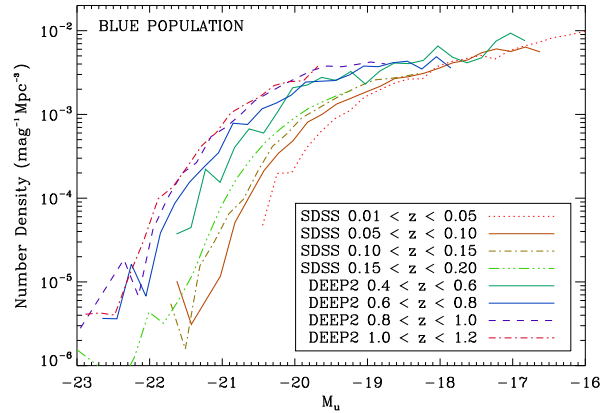}
\caption{Evolution of the {\it u}-band luminosity function for the blue population of galaxies. 0.2 mag binning is used throughout.}
\label{LFcompblue}
\end{figure}
 
\section{Results}

\subsection{Luminosity functions}

The luminosity function (LF) is defined as the number of galaxies per unit
volume per unit luminosity interval ${\rm d}l$, or alternatively magnitude
interval, ${\rm d}M$. In this paper we calculate LFs by the $1/V_{max}$
method, where $V_{max}$ is defined above. This method is non-parametric and
does not require any assumptions on the shape of the LF. It can be biased
however if there are strong density inhomogeneities in the field. The
luminosity function is estimated as follows with the number of galaxies per
Mpc$^{3}$ per absolute magnitude bin given by:
\begin{equation}
 \phi(M)=\frac{1}{\Delta M}\sum_{i}^{{N}_{g}}\left(\frac{1}{{C}_{i}\:{V}_{max,i}} \right)
\end{equation}
where $C_{i}$ is the completeness for each galaxy described in the previous
section.  

LFs were produced for the red, blue and combined populations of
galaxies, for the same 8 redshift slices as the CMDs above. For the
SDSS {\it u}GS data we produced LFs for 4 redshift slices over the
range $0.01 < z < 0.2$. For the DEEP2 data we produced LFs for 4
redshift slices between $0.4 < z < 1.2$. The data were binned in {\it
  u}-band mag bins of width 0.2 mags, apart from the DEEP2 red
population, $0.4 < z < 0.6$ and $0.6 < z < 0.8$ slices, where 0.4 mag
binning was used. The properties of the redshift slices including the
numbers of galaxies in the different populations and volumes can be
seen in Table \ref{table2}. Figs. \ref{LFcomp}, \ref{LFcompred} and
\ref{LFcompblue} show the binned LFs for the 8 redshift slices plotted
together for the combined, red and blue populations of galaxies. The
LFs of all the different populations show a shift to higher
luminosities at greater redshift.

By binning the galaxies into 8 redshift slices it ensures that
  there is reasonable resolution of the evolution ($\sim$ 10 per cent
  in luminosity per SDSS slice and $\sim 20$ per cent in the DEEP2
  slices). It also ensures there are sufficient galaxies to produce
  LFs with good signal to noise after dividing into the red and blue
  populations. This is something particularly important for the 
  0.4--0.6 DEEP2 slice which
  has fewer galaxies.

\begin{table*}
\caption{Spectroscopic sample properties}
\label{table2}
\begin{center}
\begin{tabular}{lccccc} \hline
Redshift Slice &Sample &$N{_{ALL}}$ &$N{_{RED}}$ &$N{_{BLUE}}$ &Volume ($10^{6}$ Mpc$^{3}$)\\
\hline 
0.01-0.05                     &SDSS   &6174  &1708  &4466  &0.263\\
0.05-0.10                     &SDSS   &14002 &4414  &9588  &1.781\\
0.10-0.15                     &SDSS   &13559 &4302  &9297  &4.614\\
0.15-0.20                     &SDSS   &7840  &1899  &5941  &8.552\\
0.40-0.60                     &DEEP2  &1955  &316   &1639  &2.184\rlap{$^a$}\\
0.60-0.80                     &DEEP2  &6914  &1296  &5618  &3.404\\
0.80-1.00                     &DEEP2  &9818  &1363  &8455  &4.488\\
1.00-1.20                     &DEEP2  &5874  &616   &5258  &5.374\\                          
\hline
\end{tabular}
\end{center}
$^a$This volume applies to all 4 DEEP2 fields, although most galaxies
in this redshift slice are contained within the EGS.
\end{table*}    

For an analysis of the evolution of the LFs, Schechter functions
\citep{Schechter76} were fitted to the binned data. The Schechter function is
parametrized as:
\begin{equation}
  \phi(L)=\frac{{\phi}^*}{{L}^{*}}{e}^{-L/{L}^{*}}\left( {\frac{L}{{L}^{*}}} \right)^\alpha
\end{equation}
where: $L^{*}$ ($M^{*}$) is the characteristic luminosity (magnitude) of a
galaxy at the `knee' of the function; $\phi^{*}$ is the number density of
galaxies around $L^{*}$ and $\alpha$ is the faint-end slope of the function.

Schechter functions are fitted to the LFs using a least-squares fitting method
to find the parameters $\phi$, $M^{*}$ and $\alpha$. In order to quantify the
evolution we choose to fix the faint-end slope of the LFs for the different
populations on the assumption that it remains constant with redshift. This is
due to the faint-end slope being poorly sampled in higher redshift slices. The
value we take as the faint-end slope is the one that minimises the total
${\chi}^{2}$ over all redshift slices. After doing this we find best fitting
faint-end slopes of $\alpha = -1.3$ for the blue population, $\alpha = -0.35$
for the red population and $\alpha = -1.0$ for the combined LFs.

Using a slope of $\alpha = -1.0$ for the combined population LFs is consistent
with the results of \citet{Blanton03} who used a sample of 22\,020 galaxies
with $u < 18.4$ and in the redshift range $0.02 < z < 0.14$ from SDSS DR2, to
obtain a slope of $\alpha = -0.92 \pm 0.07$. \citet{Baldry05} determined
$\alpha = -1.05 \pm 0.08$ for $z < 0.06$ and more recently \citet{Dorta08}
used a much larger sample of 192\,068 galaxies from SDSS DR6, with $16.45 < u
< 19.0$ and a redshift range $0.02 < z < 0.19$, to obtain a slope of $\alpha =
-1.01 \pm 0.03$.

 Using photometric redshifts from the COMBO-17 Survey
  \citet{Wolf03}, produce near-UV (280-nm) LFs for 25\,000 galaxies in
  the range $0.2 < z < 1.2$, as a function of SED type. This study
  finds no evolution in the faint-end slope for 4 different SED
  types. For reddest SED type 1 galaxies, the faint-end slope is found
  to be $\alpha = -0.5 \pm 0.2$ and for the the bluest, SED type 4 galaxies a
  slope of $\alpha = -1.5 \pm 0.06$ is found.

 \citet{Ilbert05} using first epoch data from the VIMOS-VLT Deep Survey
  (VVDS) \citep{LeFevre04}, measure an evolution in the faint end-slope of the
  {\it U}-band LF and find that it steepens from $\alpha = -1.05 \pm 0.05$ in the range
  $0.05 < z < 0.2$ to $\alpha = - 1.44 \pm 0.2$ in the range $0.8 < z < 1.0$. In our
  study we do not observe this steepening, but in fact the opposite, with $z =
  0.4$--0.8 DEEP2 redshift slices having shallower slopes than the SDSS slices
  (if fitted separately). 

  While allowing $\alpha$ to vary provides a better fit to the data,
  it should be noted that the galaxies with the faintest magnitudes
  within each slice have the largest volume and completeness
  corrections.  In particular, the faint DEEP2 data have a low
  redshift success rate and the assumption that the missed redshifts
  have the same distribution is a significant source of uncertainty
  (cf.\ the `average' model of \citealt{Faber07}).
  The focus of this paper is on extracting reliable luminosity
  densities, and evolution in $M^{*}$ is more robustly determined
  using fixed $\alpha$ given the well-known degeneracy in the
  $M^{*}$-$\alpha$ plane.

\begin{figure*}
\includegraphics[width=0.9\textwidth]{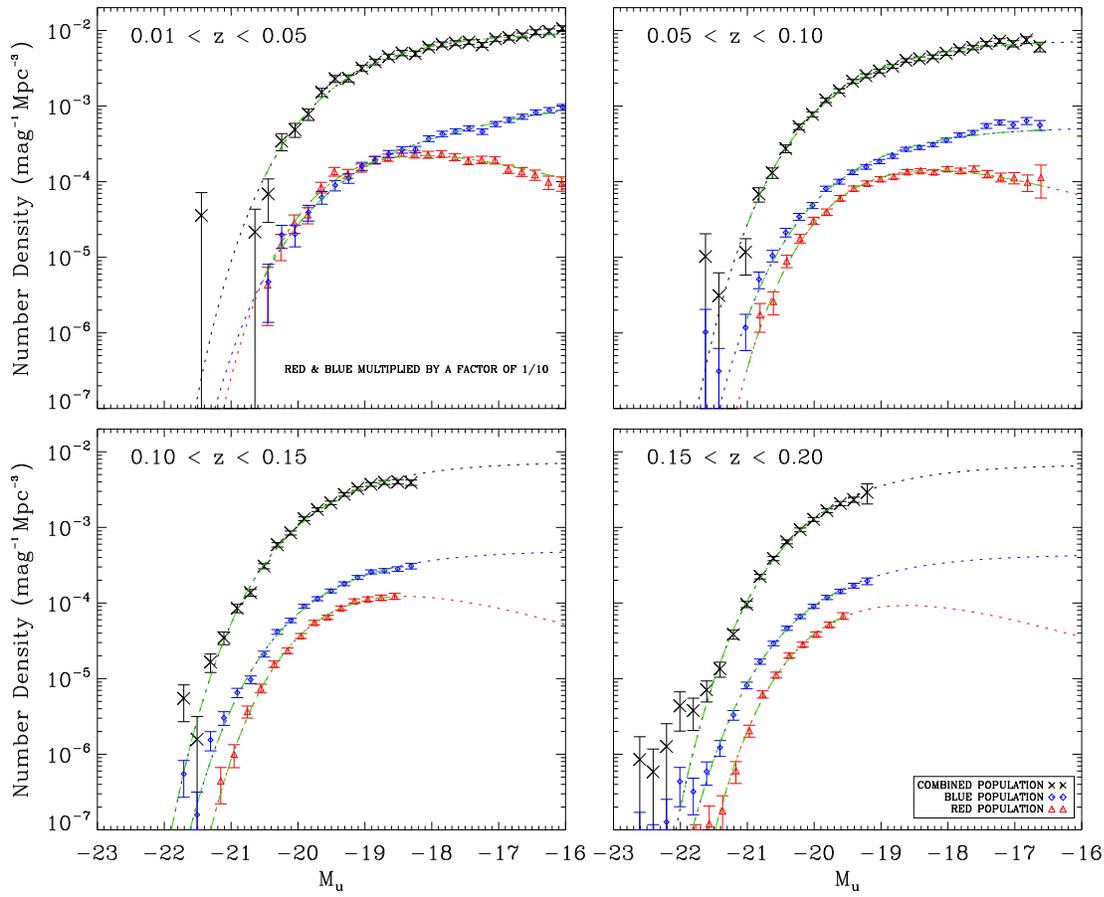}
\caption{Luminosity functions of the four SDSS {\it u}GS redshift slices, for the red, blue and combined populations of galaxies, produced using the $1/V_{max}$ method. The dotted lines show an extrapolation of the Schechter function beyond the fitted magnitude range. The LFs for the red and blue populations are scaled down by a factor of 10. Errors on the data points are Poisson errors added in quadrature to an error of 5 per cent of the number density, to take into account systematic uncertainties in the binning process. $\alpha$ is fixed at $-0.35$, $-1.3$ and $-1.0$ for the red, blue and combined populations respectively.}
\label{SDSSallLFs}
\end{figure*}

\begin{figure*}
\includegraphics[width=0.9\textwidth]{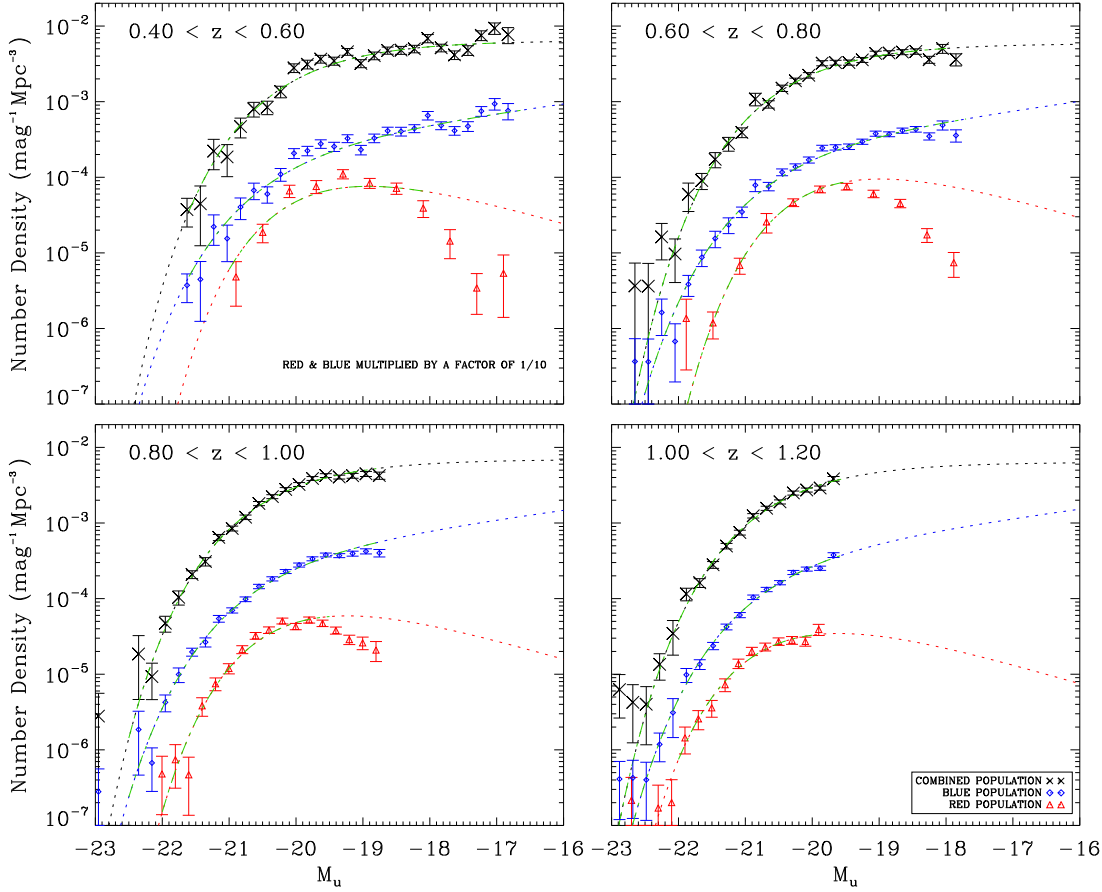}
\caption{Luminosity functions for the four DEEP2 redshift slices, for the red, blue and combined populations of galaxies. For details see Fig.~\ref{SDSSallLFs}. Note the deficit of faint red galaxies below the fit for three redshift slices because $\alpha$ is constrained by the lowest two SDSS slices.} 
\label{DEEP2allLFs}
\end{figure*}

The Schechter function fits for the red, blue and combined populations can be
seen in Figs.~\ref{SDSSallLFs} and ~\ref{DEEP2allLFs}. Errors on the number
density used in the fitting process and shown in the figures are Poisson
errors added with an error of 5 per cent of the number density, in quadrature,
to take into account systematic uncertainties in the binning process. Dotted
lines in the figures represent the regions outside the magnitude ranges used
to fit the Schechter function. It is notable that while a constant value of
$\alpha$ is suitable to describe the evolution of the blue and combined
population LFs, there is a deficit of faint red galaxies below the Schechter
fit in the higher-redshift DEEP2 data (cf.\ evolution of the field
dwarf-to-giant ratio, \citealt{Gilbank08}).  The main focus of this paper,
however, is on the blue sequence and its relationship to cosmic SFR
evolution. Tables \ref{table3}, \ref{table4} and \ref{table5} display how the
fitted Schechter parameters evolve along with the luminosity density
(described in the next subsection) from an integration of the LF
($\rho_{FIT}$) and from a summation of the luminosities of all the galaxies
($\rho_{SUM}$).
 
\begin{table*}
\caption{Schechter function parameters for all galaxies with an assumed $\alpha = -1.0$.}
\label{table3}
\begin{center}
\begin{tabular}{cccccc} \hline
$<z>$ &$N{_{GAL}}$ &$M{^{*}}$ &$\phi{^{*}}$ ($10^{-4}$ gal Mpc$^{-3}$) & $\log\rho_{FIT}$ (W Hz$^{-1}$ Mpc$^{-3}$) & $\log\rho_{SUM}$ (W Hz$^{-1}$ Mpc$^{-3}$) \\
\hline 
0.030 & 6174  & --18.89  $\pm$ 0.03 &103.68 $\pm$ 2.55 & 19.21 $\pm$ 0.01 & 19.23 $\pm$ 0.01 \\
0.075 & 14002 & --19.12  $\pm$ 0.02 & 81.60 $\pm$ 1.79 & 19.19 $\pm$ 0.01 & 19.17 $\pm$ 0.05 \\
0.125 & 13599 & --19.27  $\pm$ 0.02 & 80.57 $\pm$ 2.55 & 19.25 $\pm$ 0.01 & 19.10 $\pm$ 0.03 \\
0.175 & 7840  & --19.44  $\pm$ 0.02 & 73.96 $\pm$ 3.78 & 19.28 $\pm$ 0.02 & 19.00 $\pm$ 0.05 \\
0.500 & 1955  & --19.80  $\pm$ 0.05 & 69.71 $\pm$ 2.74 & 19.41 $\pm$ 0.02 & 19.42 $\pm$ 0.05 \\
0.700 & 6914  & --20.07  $\pm$ 0.03 & 63.71 $\pm$ 1.99 & 19.46 $\pm$ 0.01 & 19.44 $\pm$ 0.03 \\
0.900 & 9818  & --20.18  $\pm$ 0.02 & 75.32 $\pm$ 2.27 & 19.59 $\pm$ 0.01 & 19.48 $\pm$ 0.13 \\
1.100 & 5874  & --20.28  $\pm$ 0.03 & 69.17 $\pm$ 2.90 & 19.59 $\pm$ 0.02 & 19.39 $\pm$ 0.14 \\            
\hline
\end{tabular}
\end{center}
\end{table*} 

\begin{table*}
\caption{Schechter function parameters for red galaxies with an assumed $\alpha = -0.35$.}
\label{table4}
\begin{center}
\begin{tabular}{cccccc} \hline
$<z>$ &$N{_{GAL}}$ &$M{^{*}}$ &$\phi{^{*}}$ ($10^{-4}$ gal Mpc$^{-3}$) & $\log\rho_{FIT}$ (W Hz$^{-1}$ Mpc$^{-3}$) & $\log\rho_{SUM}$ (W Hz$^{-1}$ Mpc$^{-3}$)\\
\hline 
0.030 & 1708 & --18.60  $\pm$ 0.03 & 60.91 $\pm$ 1.74 & 18.82 $\pm$ 0.01 & 18.82 $\pm$ 0.12 \\
0.075 & 4414 & --18.69  $\pm$ 0.02 & 38.80 $\pm$ 0.93 & 18.66 $\pm$ 0.01 & 18.65 $\pm$ 0.07 \\
0.125 & 4302 & --18.85  $\pm$ 0.02 & 33.66 $\pm$ 1.21 & 18.67 $\pm$ 0.02 & 18.57 $\pm$ 0.03 \\
0.175 & 1899 & --19.07  $\pm$ 0.03 & 25.56 $\pm$ 1.71 & 18.62 $\pm$ 0.03 & 18.42 $\pm$ 0.08 \\
0.500 & 316  & --19.37  $\pm$ 0.08 & 20.85 $\pm$ 1.62 & 18.67 $\pm$ 0.03 & 18.69 $\pm$ 0.08 \\
0.700 & 1296 & --19.44  $\pm$ 0.06 & 26.06 $\pm$ 3.07 & 18.78 $\pm$ 0.05 & 18.71 $\pm$ 0.05 \\
0.900 & 1363 & --19.71  $\pm$ 0.04 & 16.28 $\pm$ 0.85 & 18.69 $\pm$ 0.02 & 18.60 $\pm$ 0.14 \\
1.100 & 616  & --20.07  $\pm$ 0.05 & 9.52 $\pm$ 0.77  & 18.60 $\pm$ 0.04 & 18.48 $\pm$ 0.14 \\
\hline
\end{tabular}
\end{center}
\end{table*} 

\begin{table*}
\caption{Schechter function parameters for blue galaxies with an assumed $\alpha = -1.3$.}
\label{table5}
\begin{center}
\begin{tabular}{cccccc} \hline
$<z>$ &$N{_{GAL}}$ &$M{^{*}}$ &$\phi{^{*}}$ ($10^{-4}$ gal Mpc$^{-3}$) & $\log\rho_{FIT}$ (W Hz$^{-1}$ Mpc$^{-3}$) & $\log\rho_{SUM}$ (W Hz$^{-1}$ Mpc$^{-3}$)\\
\hline 
0.030 & 4466 & --19.03  $\pm$ 0.05 & 44.50 $\pm$ 1.73 & 19.01 $\pm$ 0.02 & 19.01 $\pm$ 0.09\\
0.075 & 9588 & --19.09  $\pm$ 0.02 & 57.60 $\pm$ 1.47 & 19.03 $\pm$ 0.01 & 19.01 $\pm$ 0.04\\
0.125 & 9297 & --19.28  $\pm$ 0.02 & 53.73 $\pm$ 1.86 & 19.09 $\pm$ 0.02 & 18.94 $\pm$ 0.03\\
0.175 & 5941 & --19.51  $\pm$ 0.02 & 47.85 $\pm$ 2.60 & 19.12 $\pm$ 0.02 & 18.87 $\pm$ 0.05\\
0.500 & 1639 & --20.17  $\pm$ 0.06 & 32.64 $\pm$ 1.67 & 19.34 $\pm$ 0.02 & 19.33 $\pm$ 0.04\\
0.700 & 5618 & --20.37  $\pm$ 0.04 & 33.34 $\pm$ 1.24 & 19.42 $\pm$ 0.02 & 19.35 $\pm$ 0.03\\
0.900 & 8455 & --20.39  $\pm$ 0.03 & 48.23 $\pm$ 1.72 & 19.59 $\pm$ 0.02 & 19.42 $\pm$ 0.13\\
1.100 & 5258 & --20.46  $\pm$ 0.03 & 49.13 $\pm$ 2.43 & 19.62 $\pm$ 0.02 & 19.34 $\pm$ 0.14\\
\hline
\end{tabular}
\end{center}
\end{table*} 

\subsection{Luminosity density}

After fitting Schechter functions to the binned LFs, these can be used to
estimate the total comoving luminosity density (LD), with no extinction
correction, in each of the redshift slices for the red and blue
populations. We assume that the fitted Schechter function is valid outside the
fitted magnitude range.  The LD is then given in magnitudes per Mpc$^{3}$ by
\begin{equation}
j =M^{*}-2.5\log[(\phi^{*}/$Mpc$^{-3})\Gamma_{f}(\alpha+2)]
\end{equation}
where $\Gamma_{f}$ is the gamma function, and in linear units by       
\begin{equation}
{\rho}_{L} ={10}^{(34.1-j)/2.5}\; $W H{z}$^{-1} $Mpc$^{-3} 
\end{equation}
Here $j$ is in AB mag Mpc$^{-3}$.

\begin{figure}
\includegraphics[width=0.47\textwidth]{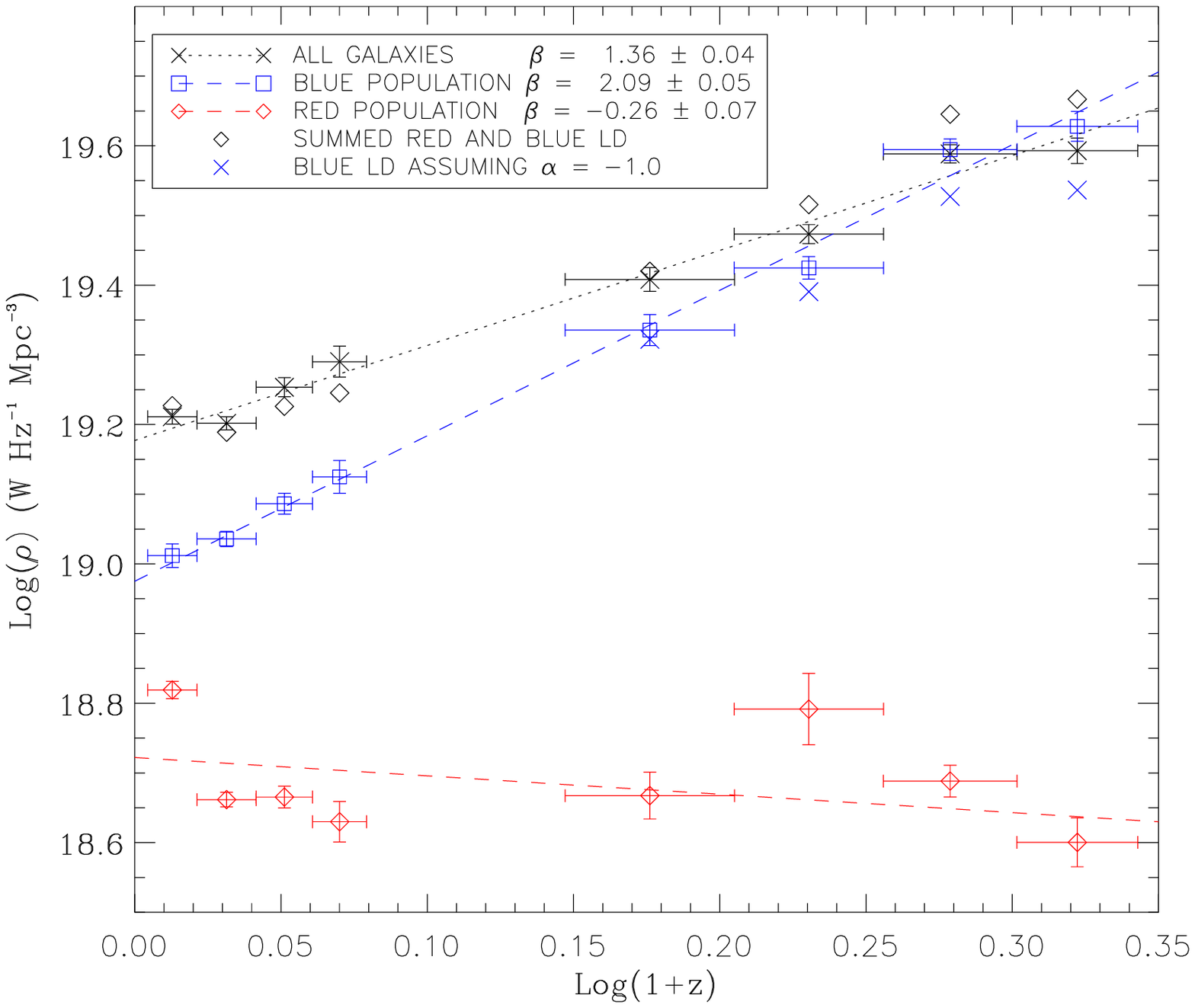}
\caption{Evolution of the {\it u}-band luminosity density for the different
  galaxy populations. The LD for the combined population is less than the LD
  for the blue population in the two highest redshift slices. This is due to
  our fixing the slope of the LF for the different populations. Black diamonds
  represent the LD obtained from the addition of red and blue LD. Vertical
  bars indicate $1\sigma$ Poisson errors on the LD. Horizontal bars represent
  the redshift range. The dashed lines represent fits (eq.~\ref{eqn:beta}) 
  to the data. Blue crosses indicate the effects of assuming
    $\alpha = -1.0$ for the blue DEEP2 population.}
\label{ALLrhoZplot}
\end{figure}

Fig.~\ref{ALLrhoZplot} shows the evolution in the integrated Schechter fit LD
for the different populations. For the the blue and combined populations, the
luminosity density has decreased steadily from redshift 1.2 to the present
day. The red population luminosity density appears to be almost constant, a
result also found by \citet{Bell04} and \citet{Faber07} who conclude this can
be explained by passive fading being compensated by a build up in the stellar
mass density. Note also that the fit to the evolution for the blue population
LD has a lower $\chi^2$ than the fit for the red population. This is not
surprising because the cosmic variance is larger for the more clustered red
population \citep{Somerville04}. In fact if the larger error bars from
  $\rho_{SUM}$ (Table~\ref{table4}) are used when fitting, the best fit shows
  a marginally rising LD with $z$ rather than declining as in the figure.

As well as calculating the LD from integration of the Schechter fit we
also estimate the LD from the summation of the galaxies' {\it u}-band
luminosity divided by $1/(V_{max}C_{i})$, in the different redshift
slices.
\begin{equation}
\rho_{SUM} = \sum_{i}^{N_{g}}\left(\frac{L_{i}}{C_{i}V_{max,i}}\right)
\end{equation}
The values of this summation for the combined, blue and red
populations can be found in Tables~\ref{table3}, \ref{table4} and
\ref{table5}. 

\begin{figure}
\includegraphics[width=0.47\textwidth]{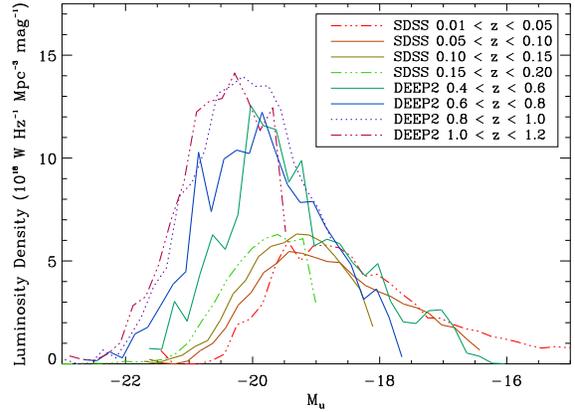}
\caption{The contribution to the luminosity density for the combined population of galaxies for the 8 redshift slices. The LD is given in units of $10^{18}$ W Hz$^{-1}$ Mpc$^{-3}$ mag$^{-1}$. 0.2 mag binning is used.}
\label{LDconALL}
\end{figure}

\begin{figure}
\includegraphics[width=0.47\textwidth]{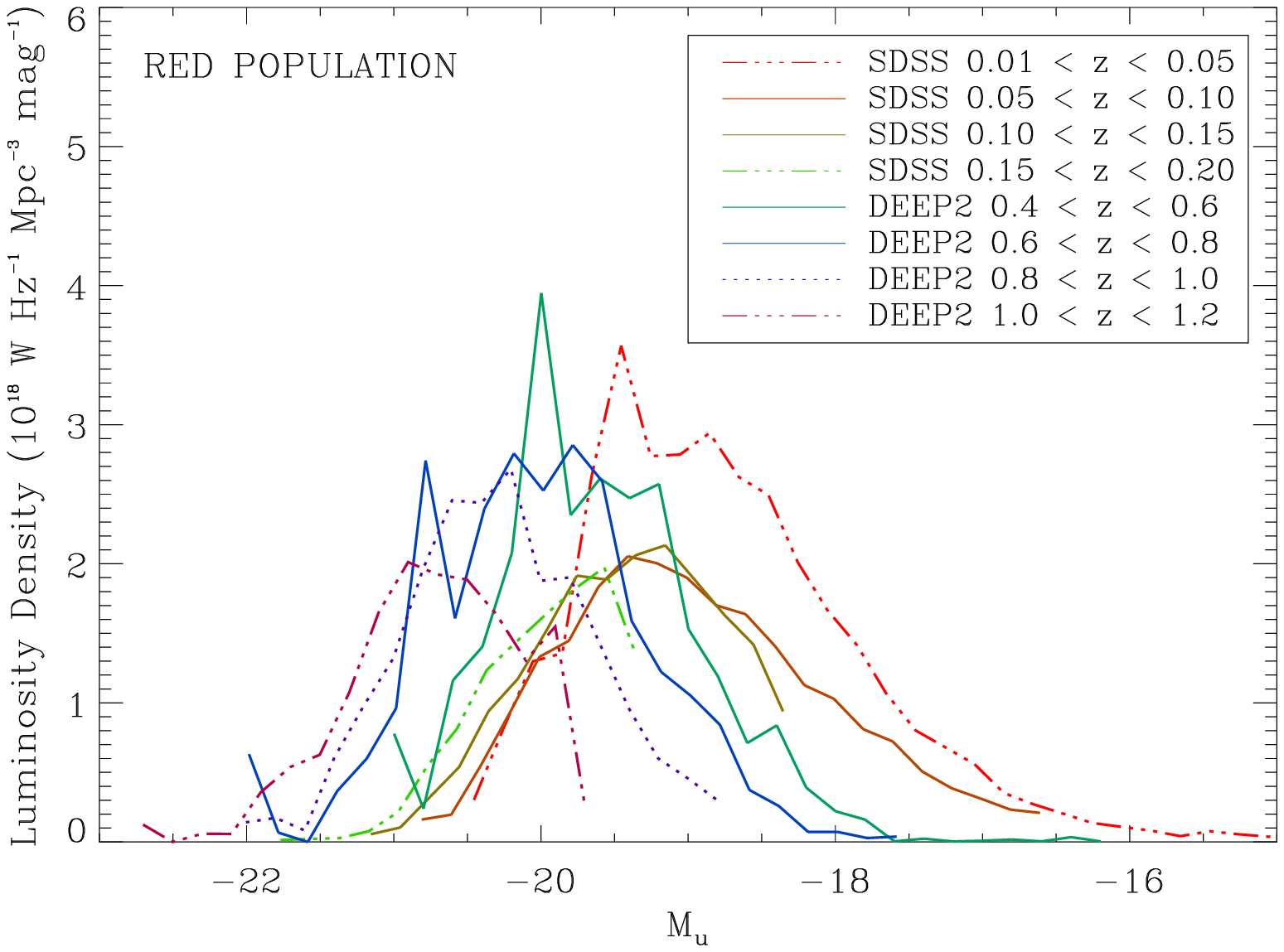}
\caption{The contribution to the luminosity density for the red population.}
\label{LDconRed}
\end{figure}

\begin{figure}
\includegraphics[width=0.47\textwidth]{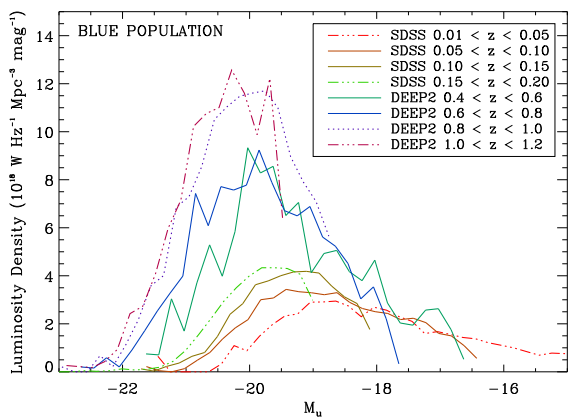}
\caption{The contribution to the luminosity density for the blue population.}
\label{LDconblue}
\end{figure}

  Errors on $\rho_{FIT}$ given in the tables are standard
  errors from the fitting procedure, while errors on $\rho_{SUM}$ were
  calculated by dividing the survey areas into sections and finding
  the standard deviation of $\rho_{SUM}$ between the sections. For the
  DEEP2 galaxies, we calculated $\rho_{SUM}$ in each of the 4 fields,
  whereas for the SDSS {\it u}GS galaxies we divided the area of the
  survey into 6 equal areas. The range in magnitudes
  $\rho_{SUM}$ applies to depends on the magnitude limits of the
  survey and is always less than the fitted magnitude range of
  $\rho_{FIT}$. As a consequence the fraction
  $\rho_{SUM}/\rho_{FIT}$ ranges from 1.05 to 0.53 for the combined
  sample of galaxies, 1.05 to 0.63 for the red population, and 1.0 to
  0.52 for the blue population. This indicates that $\rho_{FIT}$ is
  never more than a factor of 2 extrapolation from $\rho_{SUM}$. By
binning the summed LD in 0.2 mag bins and redshift slices, the
contribution of galaxies to the LD, and its evolution as a function of
absolute magnitude can be evaluated. Figs. \ref{LDconALL},
\ref{LDconRed} and \ref{LDconblue} show this evolution for all, blue
and red populations. These plots show peaks that indicate we are
sampling galaxies that contribute the most to the luminosity density.

The plots for the combined and blue populations clearly show that as
well as the total LD increasing with redshift, the galaxies that
contribute most to the LD increase in brightness with redshift. This
illustrates `downsizing', first observed by \citet{Cowie96} and also found in more
recent studies \citep{Treu05,Bundy06}, where the brighter, more
massive galaxies form their stars at higher redshift than the lower
luminosity galaxies. The LD for the red population remains more or
less constant over all redshift slices and again the galaxies that
contribute most to the LD get brighter with increasing redshift.

For the combined and blue populations we parametrize the evolution in the
integrated LD as in \citet{Lilly96} by
\begin{equation}
{\rho} _{L} \propto (1+z)^{\beta}
\label{eqn:beta}
\end{equation}
For the combined population we obtain $\beta = 1.36 \pm 0.04$ and for the blue
population we obtain $\beta = 2.09 \pm 0.05$ (assuming a constant
  faint-end slope $\alpha$). Note if we apply the $\rho_{SUM}$ errors
  before fitting, we obtain shallower slopes $\beta=1.18$ and 1.95,
  respectively, and if we apply $\alpha=-1.0$ to the DEEP2 blue population
  (Fig.~\ref{ALLrhoZplot}), the evolution is apparently lower still.  Thus, we
  estimate there is a 0.2 systematic uncertainty in the values of $\beta$
 for the combined and blue populations (larger for the red population).
In the next section we compare these values of $\beta$ with the results from
other surveys at different wavelengths.
       
\section{Discussion}

\subsection{Comparison between wavelengths}

First we compare our results with those from the {\it I}-band selected
VIMOS-VLT Deep Survey (VVDS) \citep{LeFevre04}. Using a sample of
11\,034 galaxies from the first data release of the VVDS,
\citet{Ilbert05} find that $M^{*}$ for rest-frame {\it U}-band
(360-nm) luminosity function brightens by $\sim 1.8$ mag in the
redshift range $0.05 < z < 1$.  This brightening is greater
than found in our analysis, which shows that $M^{*}$ for the combined
red and blue populations brightens by $\sim 1.4$
mag. \citet{Tresse07}, again using VVDS data, find that the evolution
in the {\it U}-band luminosity density is fitted by $\beta \approx
1.9$. By combining the LD measurements of \citet{Ilbert05} with {\it
  u}-band SDSS main sample data from \citet{Blanton03}, we find that
$\beta = 1.6 \pm 0.4$, which is greater than but consistent with our
value of $\beta = 1.36 \pm 0.2$ for the combined galaxy sample.
    
In the 150-nm far-ultraviolet (FUV) luminosity functions produced by
\citet{Arnouts05} using {\it Galaxy Evolution Explorer (GALEX)}
satellite data combined with VVDS, $M^{*}$ was determined to brighten
by 2 mag, using a sample of 1\,039 galaxies. \citet{Schiminovich05},
using the same data set, find the non-dust corrected luminosity
density to evolve with $\beta = 2.5 \pm 0.7$ out to $z\sim1$.
\citet{Wilson02} found that the 250-nm, UV LD evolves as $\beta = 1.7
\pm 1.0$ out to $z \sim 1.5$, using {\it U$^{\prime}$,B,V}
observations taken as part of a multi-passband survey of the Hubble
Deep Field and 4 other fields. Our result along with the other more
recent UV measurements rule out the early estimation of the NUV LD
evolution by \citet{Lilly96}, who found a steep evolution of $\beta
\sim 4$ using the Canada-France-Redshift-Survey (CFRS).

By going to longer wavelengths through optical to the NIR, the light
sampled in galaxies is emitted by stars of increasing lifetimes. For
the {\it B}-band (440-nm), we determined $\beta$ using LD data
published in \citet{Willmer06}. This paper made use of an earlier
DEEP2 sample of $\sim$ 11,000 galaxies. By fitting to the data we find
the LD evolves as $\beta = 0.83 \pm 0.46$. This is consistent with the
combined \citet{Blanton03} and \citet{Ilbert05} measurements, which we
find gives an evolution in the {\it B}-band of $\beta = 1.22 \pm
0.28$. These combined results also allow $\beta$ to be estimated for
the {\it V, R} and {\it I}-bands, obtaining $\beta$ values of $1.00
\pm 0.32$, $0.82 \pm 0.34$ and $0.59 \pm 0.34$ respectively.

At near-infrared (NIR) wavelengths, \citet{Pozzetti03} found from a
sample of 489 galaxies in the range $0.2 < z < 1.3$, that $\beta
\approx 0.7$ in the {\it J}-band ($1.25 \mu$m) and $\beta \approx
0.37$ in the $ K_{s}$-band ($2.17 \mu$m). This $K_{s}$-band
measurement is also consistent with more recent data from the UKIDSS
Ultra Deep Survey Early Data Release \citep{Cirasuolo07}. In this
paper LFs are produced for 6 redshift slices in the range $0.25 < z <
2.25$ using a sample of $\sim$22 000 galaxies. From the LD measured
from the first 3 redshift slices, we find that $\beta = 0.33 \pm 0.38$
in the range $0.25 < z <1.25$.

By going to still longer wavelengths in the mid-infrared, obscured star
formation can be traced from light re-emitted by dust grains. From
mid-infrared 12 $\mu$m data, LFs out to $z\sim3$ have been produced
from observations using {\it Spitzer} by \citet{Per-Gon05} for a
sample of $\sim$ 8000 galaxies. From these measurements the evolution
in LD obtained from their own form of the luminosity function out to
$z\sim1$ is best fitted by $\beta = 3.83 \pm 0.26$.  \citet{LeFloch05}
using {\it Spitzer} MIPS 24 $\mu$m data for 2600 sources, converted
into total IR luminosities find that the total IR luminosity density
evolves as $\beta = 3.9 \pm 0.4$.
 
\begin{figure}
\includegraphics[width=0.47\textwidth]{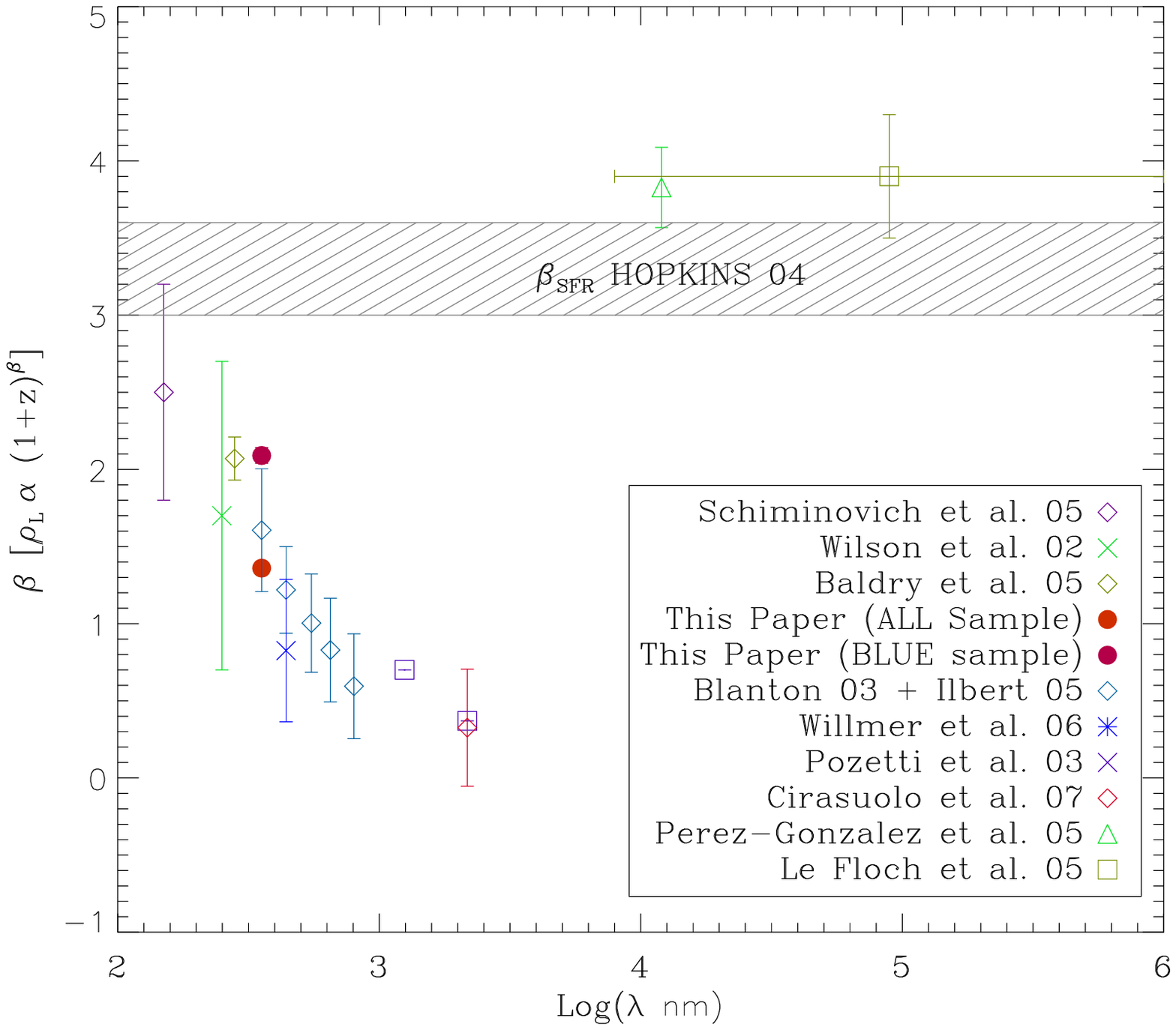}
\caption{Comparison of non dust corrected LD evolution measured at different wavelengths. Values from \citet{Schiminovich05}, \citet{Wilson02}, \citet{Baldry05} and \citet{LeFloch05} were plotted directly from the literature. \citet{Blanton03} and \citet{Ilbert05} LD results were combined and $\beta$ calculated from a fit. \citet{Pozzetti03} is an estimate quoted in their paper and has no errors. $\beta$ was calculated for \citet{Willmer06}, \citet{Per-Gon05} and \citet{Cirasuolo07} using LD values at different redshifts and fitting. The error bars for these points are $1\sigma$ errors from the fit. The shaded region indicates the evolution of the star formation rate estimated from multiwavelength data \citep{Hopkins04}. For \citet{LeFloch05} we indicate the range of wavelengths that make up the total IR.}
\label{BetaComp}
\end{figure}

Fig.~\ref{BetaComp} summarises the above results of how the luminosity
density evolution ($\beta$) varies as a function of wavelength. As wavelength
increases from the far UV to the near infrared, $\beta$ decreases, which is
because of the increasing contribution from older stellar populations to the
luminosity of the galaxies. This plot clearly shows the difference between our
results for the evolution of the {\it u}-band LD for the blue and combined
populations and the evolution of the infrared LD \citep{Per-Gon05,
  LeFloch05}. Also shown is the SFR density evolution calculated by
\citet{Hopkins04} using a compilation of X-ray, UV, $[{\rm O}_{II}]$,
H$_{\alpha}$, H$_{\beta}$, mid-IR, sub-millimetre and radio measurements
corrected for dust attenuation where necessary.

\subsection{Estimating the cosmic SFH from the {\it u}-band LD} 

Given the colour bimodality of galaxies, there is a natural separation between
blue and red populations. We can estimate the evolution in the SFR density of
the universe from the {\it u}-band LD evolution by considering just the blue
population. By doing this we effectively remove most of the contribution of
{\it u}-band luminosity produced by old stellar populations (passive red
galaxies).

In order to obtain an estimate of the evolution in the SFR density, we first
have to consider the residual {\it u}-band luminosity produced by the old
population of stars ($\tau > 1$ Gyr) in blue galaxies. To do this we perform
an analysis using PEGASE \citep{Fioc97} models to determine the contribution
of {\it u}-band luminosity from young stars with ages $\tau < 1$ Gyr old. In
these models we assume an IMF of \citet{Kroupa01}, with solar metallicity and
constant star formation rate between a formation redshift ($z_{form}$) and an
observed redshift ($z_{obs}$). For a model with $z_{form} = 6$, the fraction
of the {\it u}-band luminosity from young stars changes from 0.88 at $z_{obs}
= 1$ to 0.83 at $z_{obs} = 0$. For $z_{form} = 3$, the fraction
changes from 0.9 at $z_{obs} = 1$ to 0.84 at $z_{obs} = 0$. The assumption
that blue galaxies in general form their stars at a quasi-constant rate can be
justified considering that studies, such as \citet{Brinchmann04}, have shown
low mass galaxies to have had a roughly constant star formation rate and
\citet{James08b} find the same result for local late-type galaxies. If the
fraction of {\it u}-band luminosity produced by young stars changes by these
amounts from $z = 1$ to $z = 0$ the evolution in the LD evolution with the
contribution from old stellar populations removed is $\beta = 2.2 \pm 0.2$.

The second correction to estimate the evolution in SFR density is to consider
an increase in dust attenuation from $z = 0$ to $z = 1$, as luminosities of
characteristic galaxies at higher redshifts are brighter as found by
\citet{Cowie96} and more luminous/higher SFR galaxies have higher dust
attenuation than faint systems \citep{Hopkins01}. For a simple analysis of how
much the dust attenuation increases from $z = 0$ to $z = 1$, we use the
magnitudes of the galaxies contributing the most to the LD at $z = 0$ and $z =
1$. Using Fig.~\ref{LDconblue}, it can be seen that the main contribution to the
luminosity density at $z=1$ peaks at $M_{u} \sim -20.5$ and at $z = 0$ it
peaks at $M_{u} \sim -19$. Converting these typical galaxy magnitudes into
flux, we determine the star formation rate (in ${\rm\ M}_\odot$ yr$^{-1}$) of
the galaxies, uncorrected for dust, using the equation of \citet{Hopkins01},
which is a conversion taken from \citet{Cram98} multiplied by a factor of 5.5
to account for stars with masses $\le 5 {\rm\ M}_\odot$.

\begin{equation}
SFR _{UV} = L_{UV}\,/\,7.14 \times 10 ^{20}\, {\rm W Hz^{-1}}
\end{equation}

Dust corrected star formation rates are obtained from eq. 7 in
\citet{Hopkins01} using the SFRs of the typical galaxies. The
difference between the corrected and uncorrected SFRs gives an
effective dust attenuation of $A_{u} \sim 1.0$ mag at $z = 0$ and
$A_{u} \sim 1.25$ at $z =1$ which equates to photon escape fraction of
$\sim 0.40$ at $z =0$ and $\sim 0.32$ at $z = 1$. This is comparable
to the results of \citet{Driver08} who find the photon escape fraction
to be $\sim 0.45$ using the local galaxies.

This is reasonable as other UV studies such as \citet{Buat05} find the mean
extinction of a {\it GALEX} sample of galaxies in the local universe to be $A
= 1.1$ in the NUV (230-nm) and $A = 1.6$ for the 150-nm FUV. The dust
corrected evolution in the LD for the blue galaxies is essentially the
evolution in the SFR density of the universe excluding dusty star forming 
{\it red}
galaxies and possibly the most extreme objects with hidden star formation such
as luminous infrared galaxies (LIRGS). Assuming there is an increase in the
photon escape fraction of $0.32\pm0.02$ to $0.40\pm0.02$ from $z = 1$ to $z =
0$, this implies that $\beta_{SFR} = 2.5 \pm 0.3$.

\subsection{Further correction for dusty star forming galaxies}

After correcting for {\it u}-band LD evolution for dust and the
residual {\it u}-band luminosity from the old stellar populations, the
discrepancy still remains between our estimate in the SFR evolution
($\beta_{SFR} = 2.5 \pm 0.3$) and the results of \citet{Hopkins04}
($\beta_{SFR} = 3.3 \pm 0.3$), \citet{LeFloch05} ($\beta_{SFR} = 3.9
\pm 0.4$) and the evolution in SFR densities we determined using the
combination of UV and far IR from \citet{Takeuchi05} to give
$\beta_{SFR} = 3.4 \pm 0.6$. Possible reasons for the discrepancy
include: (i) star formation on the red sequence; (ii) a higher dust
attenuation correction for dusty star forming galaxies; (iii)
evolution of the IMF.

The first way to resolve the discrepancy would be to take into account the
star formation occurring in dusty galaxies on the red sequence. Studies such
as \citet{Franzetti07} find, using first epoch VVDS data, that the fraction of
red galaxies ($U-V>1$) that can be spectroscopically classified as star
forming galaxies with large $[O_{II}]$ equivalent widths increases with
redshift from $\sim 10$ per cent at $z = 0.5$ to $\sim 40$ at $z = 1.1$, with
35 per cent of red galaxies being spectroscopically classified as star forming
overall. Correcting for these red star-forming galaxies however would only
change evolution in the SFR by a small amount because the contribution to the
{\it u}-band LD from the red population at $z = 1$ is only of $\sim10$
per cent, as seen in Fig.~\ref{ALLrhoZplot}.

Another possibility would be to take into account the dust attenuation of
starbursting, luminous infrared galaxies (LIRGS), for which our method for
dust correction may be inadequate. \citet{Per-Gon05} find that the
contribution from LIRGS to the SFR density rises from $\sim20$ per cent at $z
= 0$ to become the dominant source of star formation at $z = 1$. In order to
obtain an evolution of $\beta_{SFR} > 3.2$, the photon escape fraction in the
$u$-band would have to drop from 0.4 at redshift $z=0$ to $< 0.2$ at
$z = 1$. However, the fossil record analysis of \citet{Panter07} implies
$\beta_{SFR}=1.5\pm0.6$ (from the first four data points of their table~3),
which is inconsistent with $\beta > 3$ ($2.5\sigma$).

A final possible way to explain the difference would be to consider that
starbursting galaxies form stars with a top-heavy IMF, which has been
suggested by \citet{Baugh05} to explain the number counts of faint
sub-millimetre galaxies, to explain the metallicities of intergalactic gas in
clusters \citep{Nagashima05a} and stars in ellipticals \citep{Nag05b}. More
recently \citet{Lacey08} used a top-heavy IMF to reproduce the evolution in
the mid-IR luminosity function observed by {\it Spitzer}. An evolving IMF has
been suggested by \citet{Wilkins08} to explain differences between
instantaneous SFRs and SFRs inferred from the stellar mass
density. \citet{Dave08} also proposes an IMF which has more high mass stars
compared to low mass at earlier epochs in order to explain differences between
the observed stellar mass-SFR relation and theoretical models. It may be
possible that the SFR evolution measured from the infrared and far UV is more
biased to the formation of the most massive stars whereas the {\it u}-band
traces star formation of a larger mass range of stars in galaxies with a
`normal' IMF. 

\section{Conclusions}

We have produced and analysed {\it u}-band luminosity functions, to quantify the evolution of the {\it u}-band luminosity density in the range $0 < z < 1.2$, for samples of galaxies taken from the SDSS {\it u}GS and DEEP2 galaxy surveys. Separating the red and blue populations of galaxies using CMDs, our main results are as follows: 

\begin{enumerate}
\item We fit Schechter functions for the red, blue and combined populations of
  the galaxies assuming constant faint-end slopes of $\alpha = -0.35$, $\alpha
  = -1.3$ and $\alpha = -1.0$, respectively.  A constant $\alpha$ with
  redshift gives reasonable fits to the blue and combined populations (this
  does not exclude some evolution in $\alpha$) but is a poor fit to the
  red sequence at $z>0.4$, which reflects significant evolution in the 
  dwarf-to-giant ratio \citep{Gilbank08}.

\item $M^{*}$ has brightened by $\sim 1.4$ mags for the combined population
  from $z=0$ to 1.2, which is similar to the findings of \citet{Ilbert05}
  using the VVDS sample who find galaxies brighten by 1.6 to 2 mags.

\item The galaxies in all populations that contribute the most to the {\it
    u}-band luminosity are observed to increase in brightness with
  redshift. This illustrates `downsizing' as seen by \citet{Cowie96} and later
  studies.

\item By parametrizing the evolution in {\it u}-band luminosity density as
  $\rho \propto (1+z)^{\beta}$, we find that the combined population of
  galaxies evolves with $\beta = 1.36\pm0.2$ and for the blue population
  $\beta = 2.09\pm0.2$. The red population LD remains approximately constant
  over time.

\item Considering just the blue population, removal of the {\it u}-band
  luminosity contribution from old stellar populations, estimated from
  population synthesis models, increases the {\it u}-band LD evolution to
  $\beta = 2.2$.  This represents the SFR density evolution uncorrected for
  dust.

\item We estimate that the average dust attenuation is 1.0 mag at $z = 0$ and
  1.25 mag at $z = 1$. By correcting the {\it u}-band LD for this extinction,
  we obtain the evolution in the SFR excluding red star-forming galaxies and
  LIRGS to be $\beta_{SFR} = 2.5\pm0.3$. This modest correction for evolution
  in dust attenuation may be appropriate for estimating the build-up of
  stellar mass, whereas the more severe dust evolution suggested by mid and
  far-IR measurements may be biased toward top-heavy IMF star formation.
\end{enumerate}

\section {Acknowledgements}

MP acknowledges STFC for a postgraduate studentship. IKB and PAJ acknowledge
STFC for funding. We acknowledge the IDL Astronomy User's Library, and IDL
code maintained by D.~Schlegel (IDLUTILS) as valuable resources.

We thank the anonymous referee for helpful suggestions, which improved
the content, clarity and presentation of this paper.
We also thank Mike Cooper of the DEEP2 collaboration for quick
responses to questions. 

Funding for the creation and distribution of the SDSS Archive has been
provided by the Alfred P.~Sloan Foundation, the Participating Institutions,
the National Aeronautics and Space Administration, the National Science
Foundation, the US Department of Energy, the Japanese Monbukagakusho and the
Max Plank Society. The SDSS web site is http://www.sdss.org.

The DEEP2 Redshift Survey has been made possible through the dedicated efforts
of the DEIMOS instrument team at the University of California, Santa Cruz and
staff at the Keck Observatory.

\bsp

\label{lastpage}

\end{document}